\documentclass[twocolumn]{aastex631}

\usepackage{amsmath, nccmath}
\usepackage{natbib}
\usepackage{multirow}
\usepackage{float}
\usepackage{ulem}

\received{September 28, 2021}
\revised{November 28, 2021}
\accepted{November 30, 2021}

\submitjournal{ApJ}

\shorttitle{Effects of Winds on the Optical Properties of SN Ib/Ic Progenitors}
\shortauthors{Jung, Yoon \& Kim}

\graphicspath{{./}{figures/}}

\begin{document}

\title{Effects of Winds on the Optical Properties of Type Ib and Ic Supernova Progenitors}

\author[0000-0002-8074-6616]{Moo-Keon Jung}
\affiliation{Department of Physics and Astronomy, Seoul National University, Seoul 08826, Korea}

\author[0000-0002-5847-8096]{Sung-Chul Yoon}
\affiliation{Department of Physics and Astronomy, Seoul National University, Seoul 08826, Korea}
\affiliation{SNU Astronomy Research Center, Seoul National University, Seoul 08826, Korea}

\author[0000-0001-9263-3275]{Hyun-Jeong Kim}
\affiliation{Korea Astronomy and Space Science Institute, Daejeon 34055, Korea}

\defcitealias{Yoon2017b}{Y17}
\defcitealias{Yoon2019}{Y19}
\defcitealias{Nugis2000}{NL}

\begin{abstract}

We investigate the effects of winds on the observational properties of Type Ib
and Ic supernova (SN Ib/Ic) progenitors using spectral models constructed with
the non-LTE stellar atmospheric code CMFGEN.  We consider SN Ib/Ic progenitor
models of the final mass range of 2.16 -- 9.09~$M_\odot$ having different
surface temperatures and chemical compositions, and calculate the resulting
spectra for various wind mass-loss rates and  wind terminal velocities.  We
find that the progenitors having an optically thick wind would become brighter
in the optical for a higher mass-loss rate (or a lower wind terminal velocity),  
because of the formation of the photosphere in the extended wind matter and the contribution from
free-free and line emissions from the wind.  As a result, for the
standard Wolf-Rayet wind mass-loss rate, helium-deficient compact SN Ic
progenitors would be  brighter in the optical by $\sim$3 mag compared to the
case without the wind effects.  
We also find that the color dependence on the photospheric temperature is non-monotonic
because of the wind effects. 
Our results imply that inferring the progenitor mass, bolometric luminosity
and effective temperature from the optical observation using the standard
stellar evolution model prediction  can be misleading.  By
comparing our fiducial model predictions with the detection limits of the
previous SN Ib/Ic progenitor searches, we conclude that a deep search with 
an optical absolute magnitude larger than $\sim -4$ is needed to directly identify most of
the ordinary SN Ib/Ic progenitors.  We discuss implications of our results for
the observed SN Ib/Ic progenitor candidates for iPTF13bvn,  SN 2019yvr and SN 
2017ein. 


\end{abstract}

\section{Introduction} \label{sec:intro}

Most supernovae (SNe) except Type Ia SNe are produced by core-collapse in massive stars at the end of their lives. The majority of core-collapse SNe belong to Type II SNe (SNe II) and their progenitors are hydrogen-rich supergiants \citep[e.g.,][]{Smith2011}. Type Ib and Ic SNe (SNe Ib/Ic) lack hydrogen lines in their spectra. This implies that their progenitors are single Wolf-Rayet (WR) stars which lost their hydrogen-rich envelopes by stellar wind mass-loss, and/or hydrogen-deficient stars produced in binary systems by interactions with their companion stars \citep[see][for a review]{Yoon2015}.  Recent observational studies indicate that the ejecta mass distribution of SNe Ib/Ic is concentrated in the range of $1-4~M_\odot$ \citep[e.g.,][]{Drout2011, Cano2013, Lyman2014, Taddia2015}. The implied progenitor masses at explosion (i.e., about $2.4 - 5.4~M_\odot$) are significantly lower than the observed WR star masses and the predictions of stellar evolutionary models of single massive stars (i.e., $M \gtrsim 8~M_\odot$; e.g., \citealt{Hamann2006, Meynet2005, Eldridge2006, Georgy2012, Yoon2015, Hamann2019, Sander2012, Sander2019}). This favors the binary star scenario for SN Ib/Ic progenitors \citep[e.g.,][]{Podsiadlowski1992a,Vanbeveren1998, Wellstein1999, Yoon2010,Eldridge2008, Yoon2015, Yoon2017a}. 

One of the best methods to study the exact nature of SN progenitors would be to directly identify the progenitor stars in the pre-explosion images. Unlike the SN II progenitors that have been observed as optically bright supergiants \citep[e.g.,][]{Smartt2009b, Smartt2015}, progenitors of SNe Ib/Ic turn out to be harder to detect. For more than 10 recently observed SNe Ib/Ic, no direct evidence for their progenitors has been found despite fairly deep observations using the \textit{Hubble Space Telescope (HST)} or other ground-based telescopes \citep[e.g.,][]{Eldridge2013, Smith2014, Smartt2015,Vandyk2017}. So far, only three cases are reported as candidates of SN Ib/Ic progenitors: iPTF13bvn \citep[Type Ib][]{Cao2013, Bersten2014, Eldridge2015}, SN 2019yvr \citep[Type  Ib][]{Kilpatrick2021}, and SN 2017ein \citep[Type Ic][]{Vandyk2018, Kilpatrick2018}, among which only the progenitor of iPTF13bvn has been conclusively identified by its disappearance \citep{Eldridge2016,Folatelli2016}. 

This implies that the majority of SN Ib/Ic progenitors are fainter in the optical than the observed WR stars in the local universe. This would be because most SN Ib/Ic progenitors are produced in binary systems and/or because SN Ib/Ic progenitor properties at the pre-SN stage is significantly different from those of observed WR stars due to the evolutionary effects \citep{Yoon2012}. Several authors make predictions on the optical properties of SN Ib/Ic progenitors using stellar evolution models of single and binary stars under the black body approximation or using detailed non-local thermodynamic equilibrium (non-LTE) stellar atmospheric models \citep{Yoon2012, Eldridge2013, Groh2013a, Groh2013b,Kim2015}. One of the important findings in these studies is that the location of the photosphere formed in an optically thick wind is an important factor to determine the optical brightness of a SN Ib/Ic progenitor \citep{Kim2015}. This means that the uncertain wind mass-loss rate at the pre-SN stage would be critical for the optical properties of SN Ib/Ic progenitors.   
 
In this study, we systematically explore the effects of winds on the optical brightness and spectra of SN Ib/Ic progenitors for a wider parameter space than considered in \citet{Kim2015}, in terms of progenitor mass, chemcial composition, wind mass-loss rate and wind terminal velocity. For this purpose we use non-LTE stellar atmospheric models calculated with the radiative transfer code CMFGEN \citep{Hillier1998a,Hillier2003}. In Section \ref{sec:model}, we explain the detailed physical assumptions and SN Ib/Ic progenitor models used in this study. In Section \ref{sec:result}, we present our calculation results and discuss observational properties resulting from the various different assumptions of the mass-loss rate and wind terminal velocity. In Section \ref{sec:implication}, we discuss  constraints on the mass-loss rate of SN Ib/Ic progenitors by comparing our models with the detection limits of undetected SN Ib/Ic progenitors provided by previous studies on pre-explosion images. We also discuss the implications of our results for the directly observed SN Ib/Ic progenitor candidates. We conclude and summarize our study in Section \ref{sec:conclusion}.

\section{Models and Physical Assumptions} \label{sec:model}

\subsection{Input Progenitor Models} \label{subsec:modelproperty}

Most of our progenitor models are based on the study of \citet[][hereafter
\citetalias{Yoon2017b}]{Yoon2017b}. These models are constructed  by
calculating the evolution of pure helium stars at solar metallicity ($Z =
0.02$) using the \texttt{BEC} code \citep[see][and its references]{Yoon2010},
until the end of core oxygen burning, for 9 different initial helium star
masses ($M_\mathrm{He,i}$ = 3.9, 4, 6, 8, 10, 12, 15, 20 and 25~$M_\odot$). The
adopted WR wind mass-loss rate is given by  the prescription of
\citetalias{Yoon2017b}.  
This prescription is based on the empirical mass-loss rates by the Potsdam
group~\citep{Hamann2006, Hainich2014} for WNe type WR stars ($\dot{M}_\mathrm{WNE}$) and by 
\citet{Tramper2016b} for WC/WO type WR stars ($\dot{M}_\mathrm{WC}$), respectively, as the following: 
\begin{equation}
    \begin{split}
        \dot{M}_\mathrm{WNE}=f_\mathrm{WR}\left(\frac{L}{L_\odot}\right)^{1.18} \left(\frac{Z_\mathrm{init}}{Z_\odot}\right)^{0.60} 10^{-11.32} \\
        \mathrm{for}\: Y=1-Z_\mathrm{init}~, 
    \end{split}
\end{equation}
\begin{equation}
    \begin{split}
     \dot{M}_\mathrm{WC}= f_\mathrm{WR} \left(\frac{L}{L_\odot}\right)^{0.83} \left(\frac{Z_\mathrm{init}}{Z_\odot}\right)^{0.25} Y^{0.85} 10^{-9.20}
       \\         \mathrm{for}\:Y<0.90~~,  
    \end{split}
\end{equation}
and 
\begin{equation}
    \begin{split}
        \dot{M}=(1-x)\dot{M}_\mathrm{WNE}+x\dot{M}_\mathrm{WC} \\
        \mathrm{for}\:0.90 \le Y<1-Z_\mathrm{init}~,
    \end{split}
\end{equation}
where 
\begin{equation}\label{eq:Mdot_Y17_x}
x=(1-Z_\mathrm{init}-Y)/(1-Z_\mathrm{init}-0.9)~. 
\end{equation}
Here the mass-loss rate is given in units of $M_\odot~\mathrm{yr^{-1}}$, and 
$Y$ and $Z_\mathrm{init}$ denote the surface helium mass fraction and the initial metallicity of WR stars, respectively.
The wind factor $f_\mathrm{WR}$ is a free parameter and  \citetalias{Yoon2017b} suggests $f_\mathrm{WR}= 1.58$ to explain the 
luminosity distribution of WR stars of different types. See 
\citetalias{Yoon2017b} for more details.  

We also take two helium-poor SN Ic progenitor models (CO2.16 \& CO3.93) from \citet[][hereafter \citetalias{Yoon2019}]{Yoon2019}, which are calculated until the pre-SN stage with the \texttt{MESA} code \citep{Paxton2011,Paxton2013,Paxton2015,Paxton2018,Paxton2019}. For these two models, the standard mass-loss rate of \citet[][hereafter \citetalias{Nugis2000}]{Nugis2000} until the core helium exhaustion and thereafter an artificially increased mass-loss rate of $\dot{M}=10^{-4}~M_\odot ~\mathrm{yr^{-1}}$ is adopted. However, because we are interested in exploring the effects of different wind properties on the resulting spectra for a given progenitor structure, the details on how each progenitor model is constructed are not important in this study. 

The physical properties of the progenitor models are summarized in Table \ref{tab:input}. The model name starting with `HE' refers to the helium-rich models, where the surface mass fraction of helium is $Y = 0.98$ and the integrated helium mass ($m_\mathrm{HE}=\int X_\mathrm{He}~dM_r$) is fairly large ($m_\mathrm{He} \ge 0.88~M_\odot$). The model name starting with `CO' refers to the helium-poor models with $Y < 0.5$ and $m_\mathrm{He} \le 0.23~M_\odot$. The numbers after HE or CO denote the progenitor masses in units of solar mass at the pre-SN phase. Recent studies indicate that not much helium can be hidden in SNe Ic spectra and here we assume that the HE and CO models represent SN Ib and Ic progenitors, respectively \citep[see][for related discussions]{Hachinger2012,Yoon2017b,Yoon2019,Dessart2020,Williamson2021}. The considered progenitor masses are $2.91 - 5.05~M_\odot$ and  $2.16 - 9.09~M_\odot$ for helium-rich and helium-poor models, respectively. This can cover most of the SN Ib and Ic progenitor mass range inferred from SN observations (i.e., the SN ejecta masses of 1.0 - 4.0~$M_\odot$) assuming that the neutron star remnant mass is about 1.4~$M_\odot$.

\begin{deluxetable*}{hlrccrrcrccccc}
\tablenum{1}
\tablecaption{Input Physical Parameters of SN Ib/Ic Progenitor Models\label{tab:input}}
\tablehead{
\nocolhead{Model} & \colhead{Name} & \colhead{$M_\mathrm{He,i}$} & \colhead{$M$} & 
\colhead{log $L$} & \colhead{$T_\star$} & \colhead{$R_\star$} &\colhead{$\dot{M}_\mathrm{fid}$} & \colhead{$v_{\infty,\mathrm{fid}}$} &
\colhead{$m_\mathrm{He}$} & \colhead{$Y$} & \colhead{log $X_\mathrm{C}$} & \colhead{log $X_\mathrm{N}$} & \colhead{log $X_\mathrm{O}$}\\ 
\nocolhead{Reference} & \colhead{} & \colhead{($M_\odot$)} & \colhead{($M_\odot$)} & \colhead{($L_\odot$)} & \colhead{(K)}  &  \colhead{($R_\odot$)}
& \colhead{($M_\odot \mathrm{\ yr^{-1}}$)} & \colhead{$\mathrm{(km\ s^{-1})}$} & \colhead{($M_\odot$)} 
& \colhead{} & \colhead{} & \colhead{}& \colhead{} 
}
\startdata
\multirow{9}{*}{\citetalias{Yoon2017b}} 
                        & HE2.91 & 3.9  & 2.91 & 4.66 & 16850  & 25.04 & 2.37e-06 & 184.49  & 1.06 & 0.98 & -3.31 & -1.82 & -3.44\\
                        & HE2.97 & 4.0  & 2.97 & 4.68 & 19060  & 19.98 & 2.51e-06 & 206.87  & 1.07 & 0.98 & -3.31 & -1.82 & -3.44\\
                        & HE4.09 & 6.0  & 4.09 & 4.97 & 36310  & 7.67 & 5.50e-06 & 335.08  & 1.11 & 0.98 & -3.66 & -1.86 & -3.09\\
                        & HE5.05 & 8.0  & 5.05 & 5.11 & 47960  & 5.22 & 8.13e-06 & 415.33  & 0.88 & 0.98 & -3.69 & -1.86 & -3.08\\
                        & CO5.18 & 10.0 & 5.18 & 5.12 & 95080  & 1.34 & 1.55e-05 & 728.36  & 0.23 & 0.42 & -0.33 & - & -1.10\\
                        & CO5.50 & 12.0 & 5.50 & 5.18 & 117300 & 0.94 & 1.26e-05 & 942.76  & 0.17 & 0.21 & -0.27 & - & -0.65\\
                        & CO6.17 & 15.0 & 6.17 & 5.22 & 133300 & 0.76 & 1.41e-05 & 1157.21 & 0.20 & 0.23 & -0.27 & - & -0.68\\
                        & CO7.50 & 20.0 & 7.50 & 5.33 & 172800 & 0.52 & 1.70e-05 & 1715.16 & 0.18 & 0.21 & -0.28 & - & -0.62\\
                        & CO9.09 & 25.0 & 9.09 & 5.43 & 191700 & 0.47 & 1.95e-05 & 2171.91 & 0.16 & 0.19 & -0.29 & - & -0.56\\
\hline
\multirow{2}{*}{\citetalias{Yoon2019}}  
                        & CO2.16 &      & 2.16 & 4.40 & 198900 &  0.13 & 3.07e-06 & 414.08  & 0.06 & 0.31 & -0.27 & - & -0.89\\
                        & CO3.93 &      & 3.93 & 4.26 & 76630  &  0.77 & 3.25e-06 & 798.00  & 0.10 & 0.49 & -0.36 & - & -1.19\\
\enddata
\tablecomments{Input model parameters of SN Ib/Ic progenitors which are given by the stellar evolution codes (\texttt{BEC} or \texttt{MESA}). Upper nine models are based on \citetalias{Yoon2017b}, and the other two models are from \citetalias{Yoon2019}. From left to right, each column has the meaning of\newline
$-M_\mathrm{He,i}$ : the initial mass of the He star at the beginning of the stellar evolution code, \newline 
$-M$ : the SN progenitor mass after finishing the stellar evolution code, \newline
$-L$ : the total bolometric luminosity of the progenitor, \newline 
$-T_\star$, $R_\star$ : the hydrostatic surface temperature and radius (without correcting the optical depth effects from the wind), \newline
$-\dot{M}_\mathrm{fid}$ : the mass-loss rate calcculated with the \citetalias{Yoon2017b} prescription 
for the given surface properties of the progenitor model,\newline
$-v_{\infty,\mathrm{fid}}$ : the wind terminal velocity calculated with the equations \ref{eq:vinf_WN} \& \ref{eq:vinf_WC},\newline
$-m_\mathrm{He}$ : the integrated helium mass,\newline
$-Y$, $X_\mathrm{C}$, $X_\mathrm{N}$, $X_\mathrm{O}$ : the surface mass fraction of helium, carbon, nitrogen and oxygen of the progenitor.
}
\end{deluxetable*}

We do not consider the rotational line broadening effect because the majority of ordinary SN Ib/Ic progenitors would be slow rotators as they lose a large amount of angular momentum via mass-loss either by winds or binary interactions \citep[see][for a detailed discussion on this issue]{Yoon2010, Yoon2015}. We also do not consider the effects of different metallicity. The metallicity can have a variety of effects on the progenitor evolution and structure,  the wind mass-loss rate, and the resulting spectra. Our preliminary results indicate that the wind density is the major factor that determines
spectral properties for a given progenitor structure and this study covers a wide range of the mass-loss rate (see Section \ref{subsec:windmodel}). Considering the full evolutionary effects
of metallicity on the progenitor evolution and structure would be a subject of future work.

Stellar evolution models predict that the suface properties (i.e., luminosity and temperature) of SN Ib/Ic progenitors do not change significantly  during the last $\sim$ 10 years before the explosion~\citep[see][]{Yoon2017a}. However, as discussed in several studies \citep[e.g.,][etc]{Fuller2018}, energy injection from the neon or oxygen burning convective region during the last evolutionary stage might affect the structure of the outermost layers of progenitor stars and induce a mass-loss enhancement. Full consideration of these effects is beyond the scope of this paper, and here we only implicitly consider possible mass-loss enhancements by assuming various mass-loss rates as explained in Section~\ref{subsec:windmodel}.

\begin{figure*}[htb]
\includegraphics[width=\textwidth]{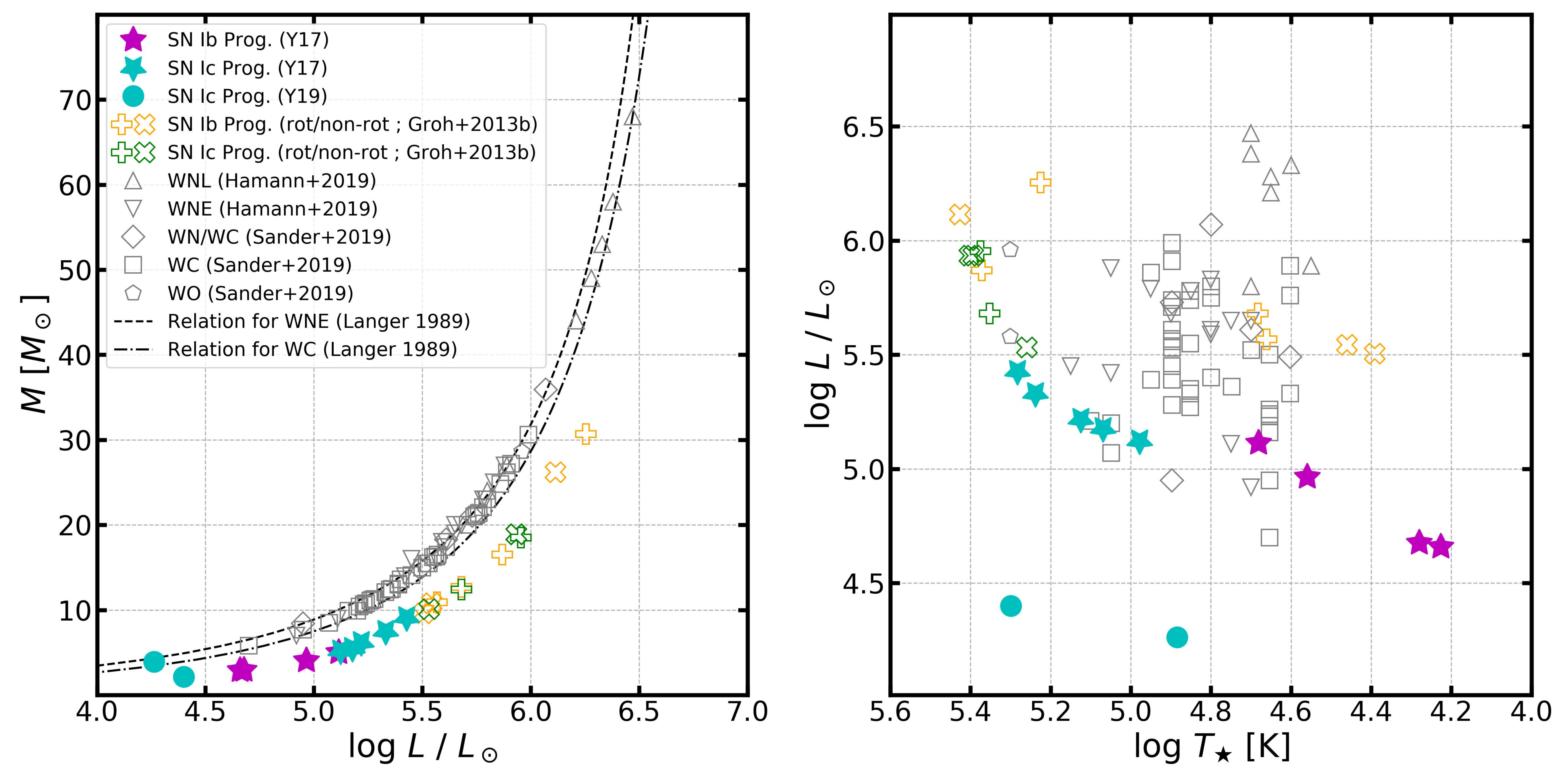}
\centering
\caption{The luminosity-mass relation (left) and the temperature-luminosity relation (right) of SN Ib/Ic progenitor models. The upright and inverted stars denote the SN Ib and Ic progenitor models of \citetalias{Yoon2017b}. The SN Ic progenitor models of \citetalias{Yoon2019} are denoted by filled circles. The rotating and non-rotating self-stripped SN Ib/Ic progenitor models of \citet{Groh2013b} are presented by  plus and X markers for comparison. The observed Galactic WR stars from \citet{Sander2019} and \citet{Hamann2019} are presented by polygons. The dashed and dashed-dotted lines show the luminosity-mass relation for WNE and WC ($Y=0.3$) stars given by \citet{Langer1989b}.\label{fig:LM}} 
\end{figure*}

Figure \ref{fig:LM} shows the luminosity-mass ($L-M$) relation and the temperature-luminosity ($T-L$) relation of our SN Ib/Ic progenitor models. More massive SN Ib/Ic progenitor models by \citet{Groh2013b} and observed Galactic WR stars \citep{Hamann2019,Sander2019} are also plotted in the figure for comparison. The comparison models from \citet{Groh2013b} are self-stripped progenitor models that have solar metallicity and initial masses of $25-120~M_\odot$ and $32-120~M_\odot$, and final masses of $10.9-30.8~M_\odot$ and $9.6-26.2~M_\odot$ for rotating and non-rotating models, respectively. These models are calculated until the end of core carbon burning with the Geneva stellar evolution code \citep{Ekstrom2012,Georgy2012} for which the WR mass-loss rates of \citetalias{Nugis2000} and \citet{Grafener2008} are used. These models are systematically more massive and luminous than our models, and the final masses are comparable to those of Galactic WR stars. The temperatures of the progenitor models in the right panel of Figure \ref{fig:LM} are the hydrostatic surface temperatures ($T_\star$) which are given by the stellar evolution models. However, the temperatures of Galactic WR stars are given by the temperatures at the location where the Rosseland mean optical depth ($\tau_\mathrm{ross}$) is 20 in stellar atmosphere models \citep{Hamann2019,Sander2019}, which might not necessarily correspond to the hydrostatic core surface temperatures.   

The observed Galactic WR stars in the left panel of Figure \ref{fig:LM} follow the theoretical $L-M$ relation for core He-burning WR stars given by \citet{Langer1989b}. This is simply because the masses of observed WR stars given by \citet{Sander2019} and \citet{Hamann2019} are based on the Langer's $L-M$ relation. However, all the progenitor models except CO3.93 are several times more luminous than the predictions of the Langer's $L-M$ relation for a given mass due to a rapid luminosity increase ($\sim0.2-0.3$ dex in logarithmic scale) at the last evolutionary stages compared to the case of helium burning stage \citep{Yoon2012, Groh2013b,Yoon2017b}. This implies that the evolutionary effects should be properly considered when inferring WR star masses from their luminosities.

\subsection{Atmospheric Modeling} \label{subsec:windmodel}

To compute our model spectra of SN Ib/Ic progenitors, we use the non-LTE atmospheric radiative transfer code CMFGEN \citep{Hillier1998a,Hillier2003}. The CMFGEN code determines the temperature distribution of the expanding atmosphere by solving the statistical and radiative equilibrium equations, and computes line and continuum formation under the assumption of the spherical symmetric geometry \citep{Hillier1990c}. 

While  the CMFGEN code computes the radiative transfer self-consistently, it does not solve the momentum equation and we need to specify the hydrodynamic structure of the wind. We assume the standard $\beta$ law for the wind velocity profile with $\beta=1$, which is given by 
\begin{fleqn}[\parindent]
\begin{equation}\label{eq:betalaw}
    \begin{split}
        v(r)=v_0+(v_\infty-v_0)\left(1-\frac{R_\star}{r}\right)^\beta~~, 
    \end{split}
\end{equation}
\end{fleqn}
where $v_0$ and $v_\infty$ denote the velocity at the hydrostatic stellar surface and the terminal velocity, respectively, and $R_\star$ means the hydrostatic stellar surface radius \citep{Lamers1999}.

We take the mass-loss rate given by \citetalias{Yoon2017b} with $f_\mathrm{WR} = 1.58$ (see above) 
as our fiducial mass-loss rate $\dot{M}_\mathrm{fid}$ for all our progenitor models (see Table \ref{tab:input}). 
We determine the wind terminal velocities of our fiducial models ($v_{\infty,\mathrm{fid}}$) by the equations for WN and WC stars from \citetalias{Nugis2000} as the following. For WN stars,
\begin{fleqn}[\parindent]
\begin{equation}\label{eq:vinf_WN}
    \begin{split}
        \mathrm{log}\: v_\infty/v_\mathrm{esc}\mathrm{(core})=\:&0.61-0.13(\pm0.09)\:\mathrm{log}\:L \\
        &+0.30(\pm0.77)\:\mathrm{log}\:Y~. 
    \end{split}
\end{equation}
\end{fleqn}
For WC stars,
\begin{fleqn}[\parindent]
\begin{equation}\label{eq:vinf_WC}
    \begin{split}
        \mathrm{log}\: v_\infty/v_\mathrm{esc}\mathrm{(core})=\:&-2.37+0.43(\pm0.13)\:\mathrm{log}\:L \\
        &-0.07(\pm0.27)\:\mathrm{log}\:Z~.
    \end{split}
\end{equation}
\end{fleqn}
The escape velocity at the stellar surface is defined as 
\begin{fleqn}[\parindent]
\begin{equation}\label{eq:v_esc}
    \begin{split}
        v_\mathrm{esc}\mathrm{(core}) =\sqrt{\frac{2GM(1-\Gamma_e)}{R_\star}}~. 
    \end{split}
\end{equation}
\end{fleqn}
The Eddington factor $\Gamma_e$ is defined as
\begin{fleqn}[\parindent]
\begin{equation}\label{eq:gamma_e}
    \begin{split}
        \Gamma_e = 7.66 \times 10^{-5} \sigma_{e} (L/L_\odot)/(M/M_\odot), 
    \end{split}
\end{equation}
\end{fleqn}
where $\sigma_{e}$ denotes the electron scattering opacity \citepalias[see][]{Nugis2000}. When calculating the wind terminal velocity, we assume the HE models as WN stars and the CO models as WC stars.  The spectra of the most massive CO model (CO9.09) might look like a WO star rather than a WC star, depending on the adopted wind parameters (see Section \ref{subsec:fiducial_models}). The observed WO stars have a very high wind terminal velocity of $\sim5000~\mathrm{km~s^{-1}}$ \citep{Drew2004a,Sander2012}. However, we consider various mass-loss rates ($\dot{M}_\mathrm{fid}\times0.1,\;0.5,\;2.0,\;5.0\;\mathrm{and}\;10.0$) and wind terminal velocities ($v_{\infty,\mathrm{fid}}\times0.1,\;0.5,\;2.0,\;\mathrm{and}\;3.0$) in our model grid, and a wind terminal velocity of $~5000~\mathrm{km~s^{-1}}$ is covered within our parameter space. 
The assumed volume filling factor $f_c$ of the wind matter, which is the inverse of the wind clumping factor, is 0.1.  
Note that WR spectra would look similar for a given value of $\dot{M}/\sqrt{f_c}v_\infty$  \citep[see][etc]{Hamann1998}. 

We take some WR and O type star models provided by the CMFGEN code as our initial trial models, which have the most similar temperature and surface gravity values to those of our input models. We first obtain the fiducial models with the fiducial mass-loss rate ($\dot{M}_\mathrm{fid}$) and the fiducial wind terminal velocity ($v_{\infty,\mathrm{fid}}$), which are determined with  $L$, $R_\star$ and chemical abundances of each model (see Table \ref{tab:input}).  Then we calculate other models with various mass-loss rates or wind terminal velocities from the fiducial models.

\section{Results of the atmospheric models} \label{sec:result}

\subsection{The fiducial models} \label{subsec:fiducial_models}

In Table \ref{tab:output}, we present the CMFGEN output results of our fiducial models. The photospheric values of the temperature and radius are obtained at the photosphere defined by the Rosseland mean opacity, which is located in the wind matter above the hydrostatic surface. The corresponding absolute optical magnitudes of different $HST$ Wide Field Planetary Camera 2 (WFPC2) filters are also given in the table. We choose the $HST$/WFPC2 filter system to compare the progenitor model magnitudes with the progenitor detection limits of previous searches (see Section \ref{subsec:detectionlimit}). Given that the instrument was replaced with the $HST$ Wide Field Camera 3 (WFC3) in 2009, we also provide the progenitor magnitudes of the $HST$/WFC3 filter system in Appendix \ref{subsec:Appendix1} to be compared with recent and future data. We find that the magntidue difference between the two filter systems is minor for HE models but significant for CO models (i.e., up to $\sim$1 mag).

\begin{deluxetable*}{hlcrcccc}
\tablenum{2}
\tablecaption{CMFGEN output parameters and optical magnitudes of SN Ib/Ic progeitors\label{tab:output}}
\tablehead{
\nocolhead{Model} & \colhead{Name} & \colhead{$T_\mathrm{eff}$} & \colhead{$R_\mathrm{phot}$} 
& \colhead{$M_{F450W}$} & \colhead{$M_{F555W}$} & \colhead{$M_{F606W}$} & \colhead{$M_{F814W}$}  \\ 
\nocolhead{Reference} & \colhead{}  & \colhead{(K)} & \colhead{($R_\odot$)} & \colhead{(mag)} & \colhead{(mag)} & \colhead{(mag)} & \colhead{(mag)} 
}
\startdata
\multirow{9}{*}{\citetalias{Yoon2017b}} 
                        & HE2.91 & 16330 & 26.65 & -5.26 & -5.19 & -5.19 & -5.16\\
                        & HE2.97 & 18390 & 21.46 & -5.11 & -5.01 & -5.00 & -4.92\\
                        & HE4.09 & 34730 & 8.38  & -4.73 & -4.54 & -4.48 & -4.31\\
                        & HE5.05 & 41910 & 6.83  & -4.56 & -4.39 & -4.34 & -4.29\\
                        & CO5.18 & 49850 & 4.88  & -4.83 & -4.84 & -4.76 & -4.74\\
                        & CO5.50 & 68200 & 2.78  & -4.54 & -4.42 & -4.24 & -4.27\\
                        & CO6.17 & 73450 & 2.51  & -4.59 & -4.46 & -4.22 & -4.22\\
                        & CO7.50 & 89770 & 1.91  & -4.53 & -4.52 & -4.25 & -4.11\\
                        & CO9.09 & 99710 & 1.73  & -4.42 & -4.39 & -4.10 & -3.97\\
\hline
\multirow{2}{*}{\citetalias{Yoon2019}}  
                        & CO2.16 & 74890 & 0.94 & -2.66 & -2.55 & -2.40 & -2.45\\
                        & CO3.93 & 48080 & 1.95 & -2.83 & -2.81 & -2.78 & -2.78\\
\enddata
\tablecomments{The meaning of each model name is same as that in Table \ref{tab:input}. From left to right, each column has the meaning of \newline
$-T_\mathrm{eff}$, $R_\mathrm{phot}$ : the effective temperature and the photospheric radius defined at the Rosseland optical depth $\tau_\mathrm{ross}$ = 2/3.\newline
$-M_{F450W,F555W,F606W,F814W}$ : the absolute magnitude in $F450W$, $F555W$, $F606W$ and $F814W$ filters in $\textit{HST}$/WFPC2.
}
\end{deluxetable*}

\begin{figure*}[htp]
\centering
\includegraphics[width=0.98\textwidth]{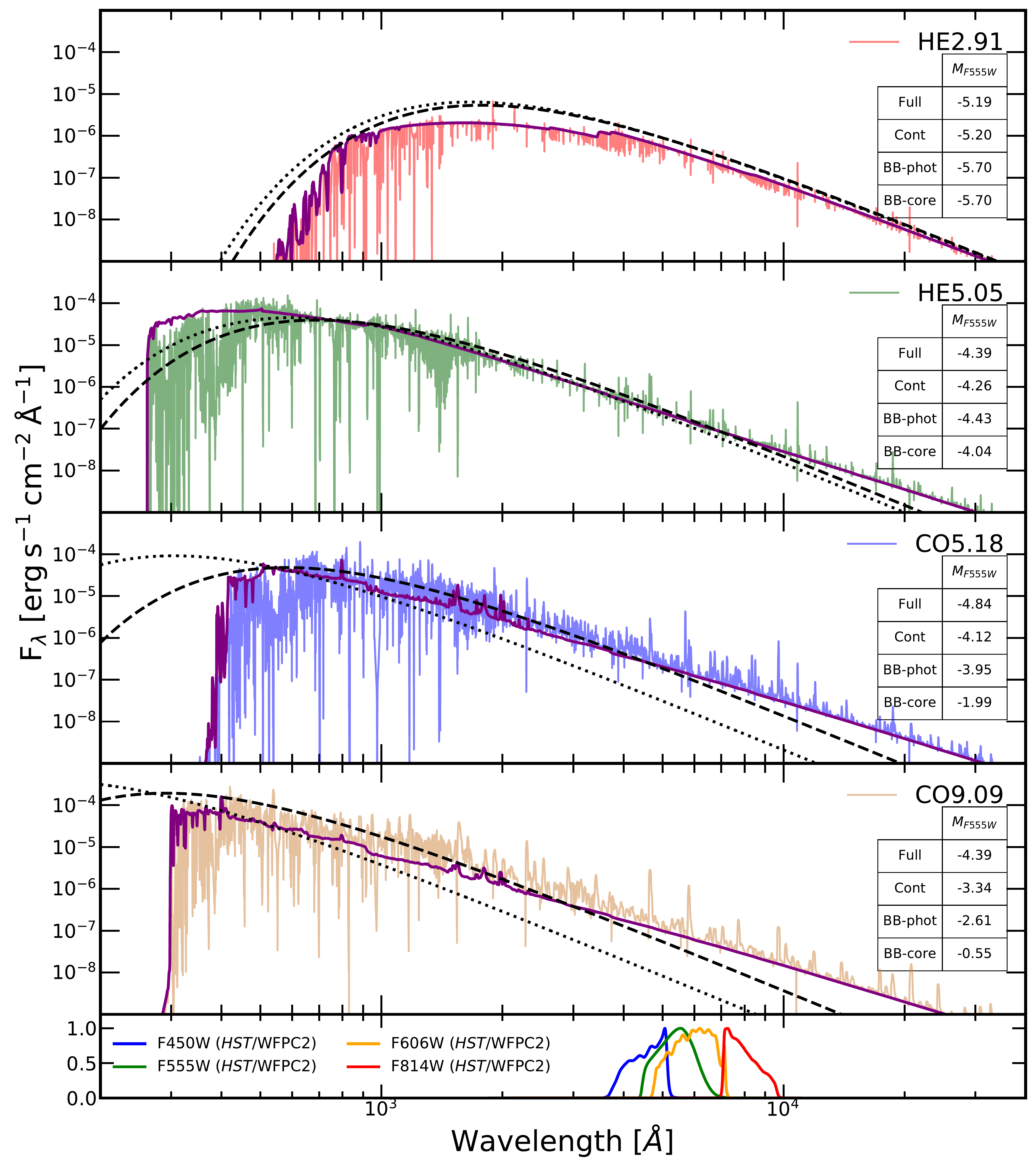}
\caption{Spectral energy distributions from different SN Ib/Ic progenitor models (HE2.91, HE5.05, CO5.18 and CO9.09). Continuum fluxes are presented by the purple line. The black body fluxes at the photosphere  (BB-phot) and the hydrostatic core (BB-core) are plotted with the dashed and dotted lines, respectively. Transmission curves of $F450W$, $F555W$, $F606W$ and $F814W$ filters in $HST$ Wide Field and Planetary Camera 2 (WFPC2) are plotted in the lowest panel. The absolute $F555W$ filter magnitude ($M_{F555W}$) of each case is presented in the inner table.  All fluxes are scaled to a distance of 10 pc.}\label{fig:sed}
\end{figure*}

We present four examples of spectral energy distributions (SEDs) in Figure \ref{fig:sed}. Here, the HE2.91 and CO9.09 progenitor  models have the lowest and highest surface temperatures among our input models from \citetalias{Yoon2017b}. The HE5.05 and CO5.18 progenitor models  have a similar mass and bolometric luminosity, but very different surface compositions and hydrostatic surface temperatures (see Table~\ref{tab:input}).

Note that a strong line blanketing in the ultra-violet wavelength region and many emission lines in the optical region are observed in the full spectra of the models. For comparison, we also plot the continuum flux given by CMFGEN, the black body flux at the photosphere (BB-phot) and the black body flux at the hydrostatic stellar surface (BB-core) of each model together. The difference between the black body flux at the photosphere and the hydrostatic stellar surface is relatively large for the CO models compared to the HE models. This is because the CO models are more compact and  have a denser wind, leading to a significant lifting-up of the photosphere compared to the HE models  (see Tables \ref{tab:input} and \ref{tab:output}). As a result, the CO models with BB-phot cases are significantly brighter in the optical than the corresponding BB-core case, and the full spectra cases are even brighter than BB-phot case (e.g., $HST$/WFPC2 $F555W$ filter magnitudes $M_{F555W} = -4.39$ and $M_{F555W} = -0.55$ for the full spectrum and
BB-core cases of CO9.09, respectively; see Figure \ref{fig:sed}), while the differences are relatively small for the HE models.

Among the fiducial models in Figure \ref{fig:sed}, the full spectrum model of HE2.91 is  fainter than the corresponding BB-core and BB-phot cases in the optical, in constrast to the other cases.  This progenitor model has a surface helium mass fraction of $Y = 0.98$ and an effective temperature of $T_\mathrm{eff}=16330$ K. The helium-rich and hydrogen-free property causes the He bound-free discontinuity at 3422~$\text{\AA}$, which corresponds to the ionization from the 2$^3$P state of \ion{He}{1}. This discontinuity is also shown in previous spectral models and observations of He-rich stars having a similar temperature range ($T_\mathrm{eff}\sim16000-24000$ K) \citep[e.g.,][]{Popper1947,Hunger1969,Rosendhal1973,Cidale2007}.  This continuous helium absorption makes the full spectrum brightness fainter than the BB-core and BB-phot brightness in the optical for HE2.91 model.  The same phenomenon is also found for the HE2.97 model that has 
a similar surface temperature to the HE2.91 model. 

In Figure \ref{fig:sed}, we find that the continuum flux exceeds the BB-phot
flux in a relatively long wavelength range for the HE5.05, CO5.18 and CO9.09
models.  For example, such an excess is found for $\lambda > 5000~\text{\AA}$ in the
CO5.18 model.  This excess is due to the free-free emission from the ionized
stellar winds, and can significantly affect optical magnitudes, especially in
red filters. 

\begin{figure}[htp]
\includegraphics[width=0.45\textwidth]{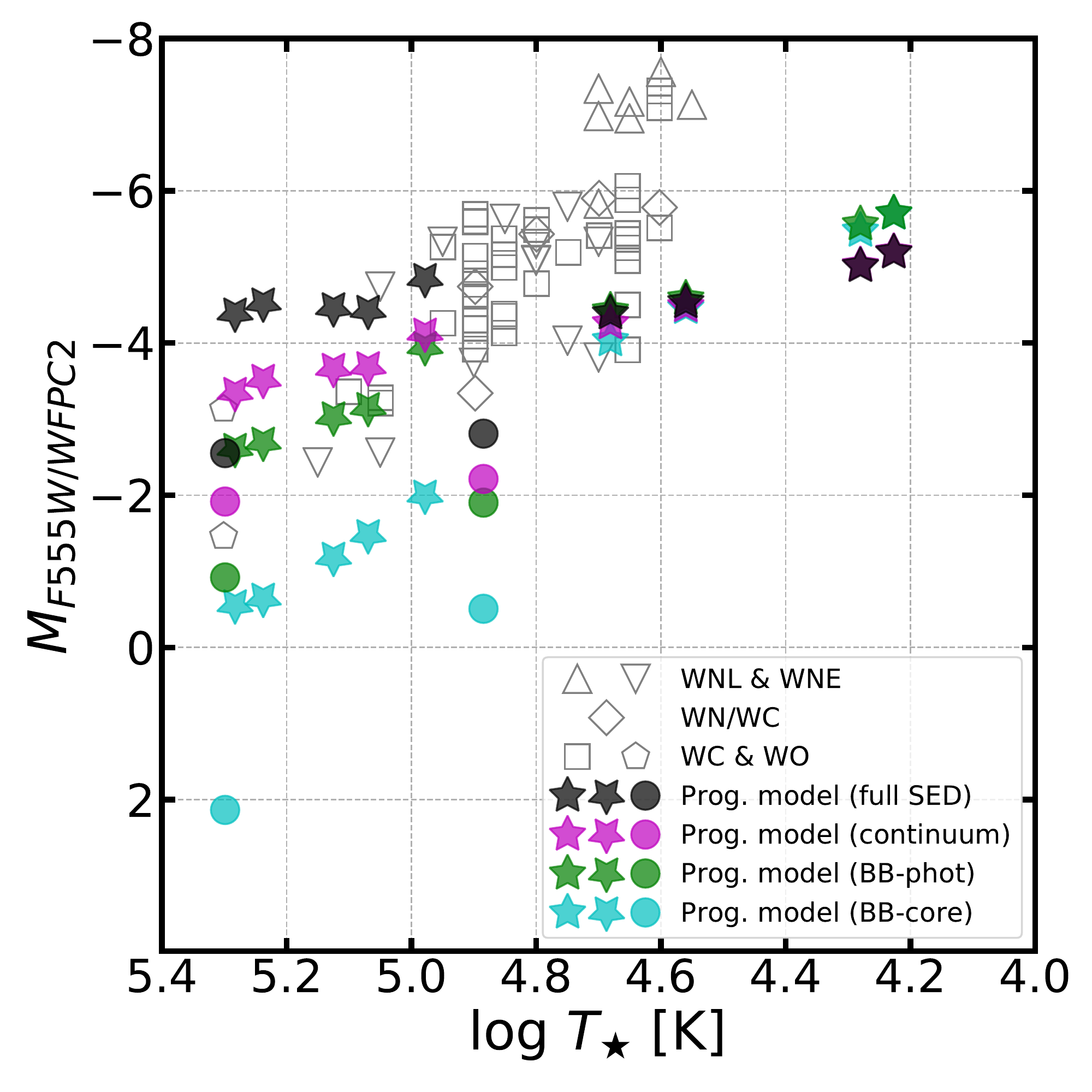}
\centering
\caption{Absolute magnitudes of SN Ib/Ic progenitors in the $F555W$ ($HST$/WPFC2) filter. The upright and inverted stars denote the HE and CO progenitor models from \citetalias{Yoon2017b}, respectively. The CO progenitor models from \citetalias{Yoon2019} are denoted by filled circles. Different colors give the optical magnitudes calculated with full SED, continuum flux, black body flux at the photosphere (BB-phot) and the hydrostatic surface (BB-core) as indicated by the labels. Galactic WR star samples from \citet{Sander2019} and \citet{Hamann2019} are plotted together for comparison (Note that the temperatures of Galacitc WR stars are calculated at the radius of $\tau_\mathrm{ross}=20$ and the absolute magnitudes of them are obtained in the Smith $v$ filter($\lambda_\mathrm{peak}=5160~\text{\AA}$ ; \citet{Smith1968a}).). \label{fig:windeffect}}
\end{figure}

In Figure \ref{fig:windeffect}, we show $M_{F555W}$ of our fiducial models calculated
with the full spectra (black markers), compared to the  continuum flux (magenta), BB-phot (green) and
BB-core cases (cyan).  It is clearly seen that for the CO models $\log
T_\star~\mathrm{[K]} > 4.8$,  the optical
brightness becomes significantly higher when the full SEDs are taken into
account than the BB-core case.  This is mainly because of the effect of
lifting-up of the photosphere with an optically thick wind, as discussed above.

On the other hand,  the magnitudes $M_{F555W}$ of the HE models $\log
T_\star~\mathrm{[K]} < 4.8$ do not change
much for different cases of full SED, continuum flux, BB-phot, and BB-core.
This is because the HE models have a relatively large radius at the hydrostatic
surface and the resulting wind density with $\dot{M}_\mathrm{fid}$ is not high
enough to cause a significant lifting-up effect of the photosphere. For the
cases of HE2.91 and HE2.97, continuum luminosity is fainter than BB-core
luminosity in the optical due to the continuous He absorption as discussed above.

\begin{figure*}[htp]
\gridline{\fig{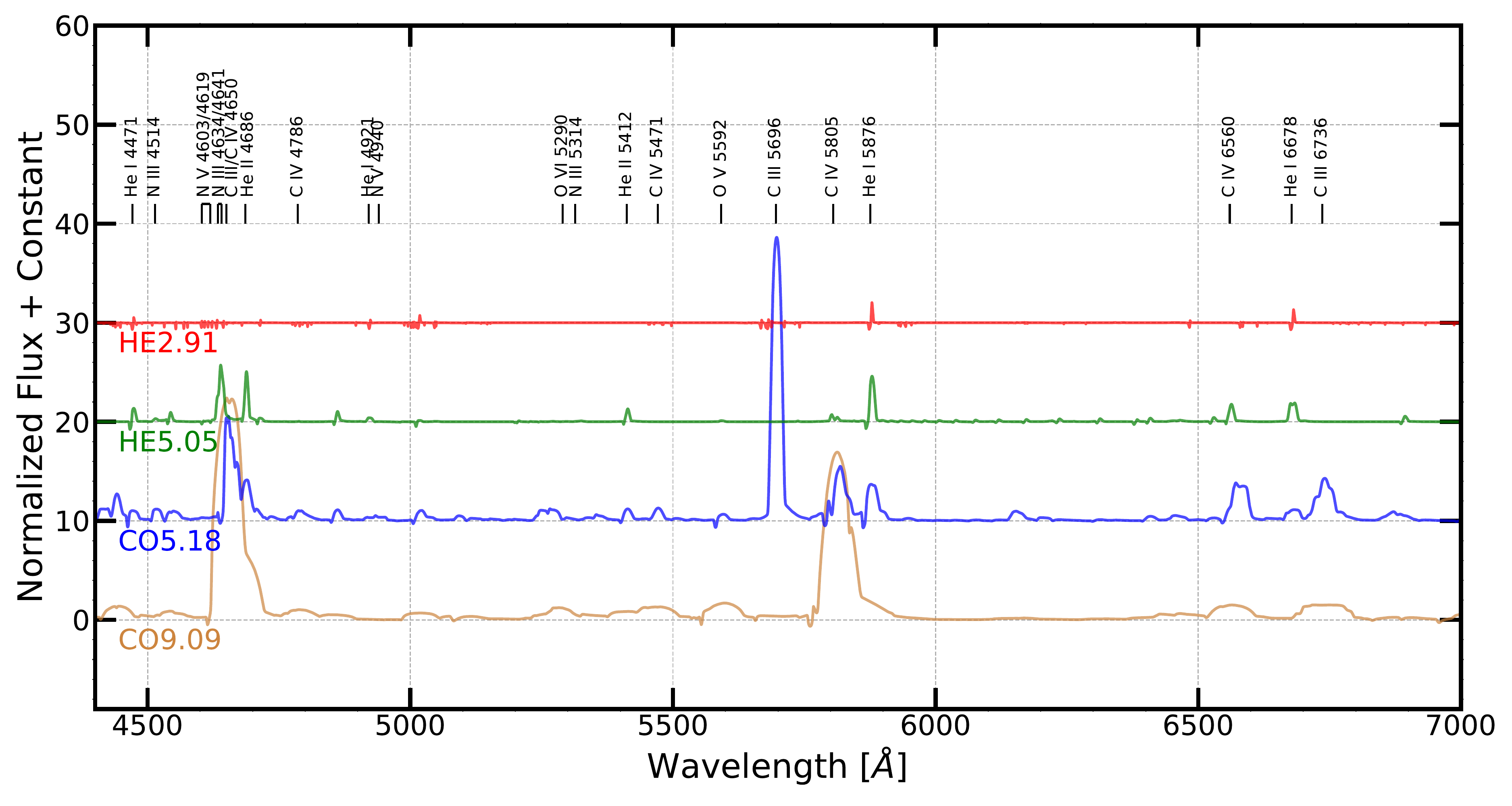}{\textwidth}{(a)}}
\gridline{\fig{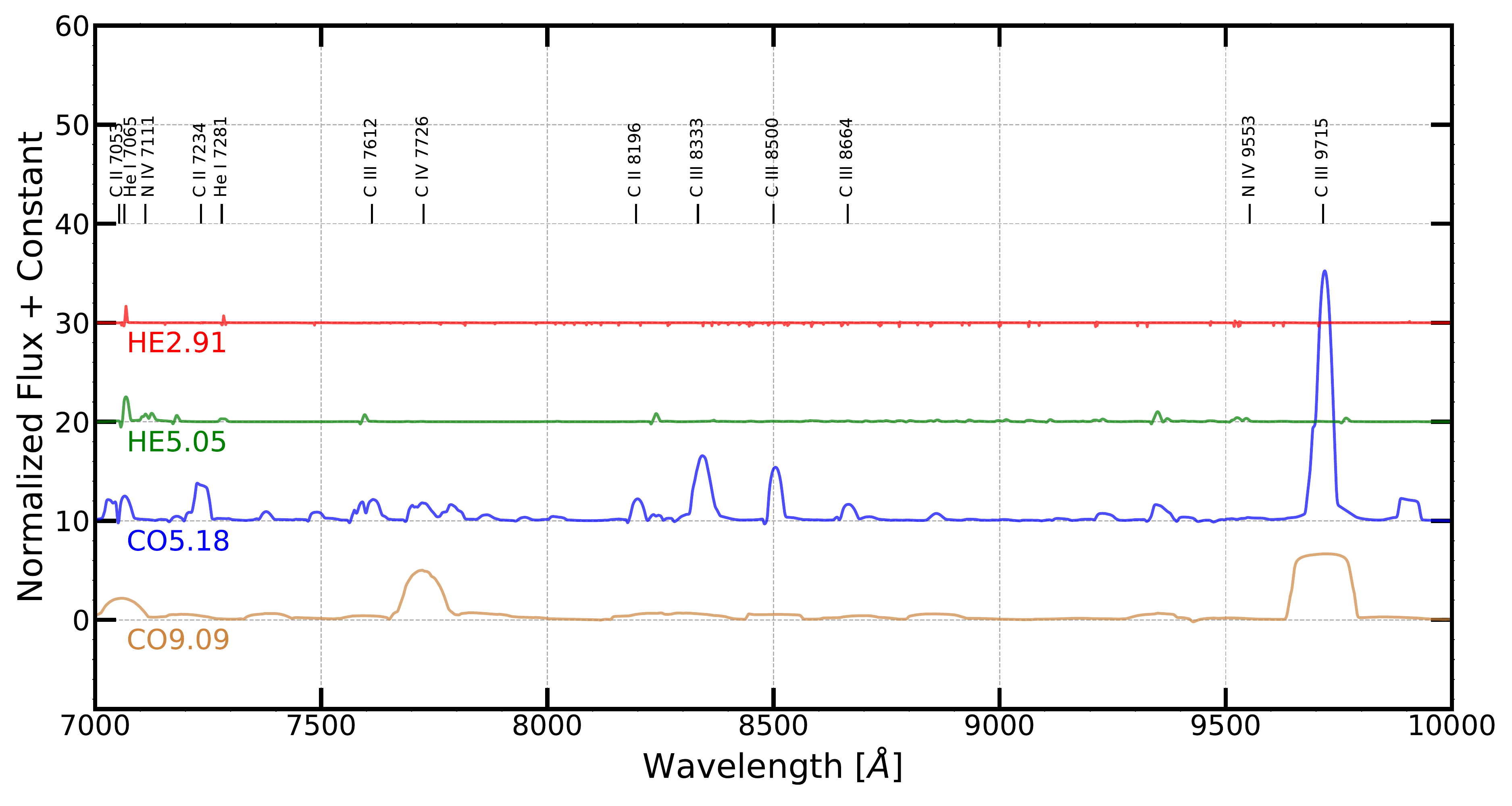}{\textwidth}{(b)}}
\caption{Normalized optical spectra of some selected SN Ib/Ic progenitor models in the wavelength
ranges (a) $4400~\text{\AA} \leqslant \lambda \leqslant 7000~\text{\AA}$ and (b) $7000~\text{\AA} \leqslant \lambda \leqslant 10000~\text{\AA}$, which are correspond to the wavelength range of $F555W$ and $F814W$ filters ($HST$/WFPC2), respectively. \label{fig:normalizedSED}} 
\end{figure*}

Figure \ref{fig:normalizedSED} shows the optical spectra normalized by continuum, 
which can be used for WR classification~\citep{Smith1968b}.
In  HE2.91,  most lines appear as absorption lines except for
a few P Cygni profiles of He I. In HE5.05, many P Cygni profiles and emission
lines are found. 
HE5.05 may be classified as WN7-8 due to its strong
\ion{N}{3} $\lambda4634$ emission line, and the similar strength between
\ion{He}{1} and \ion{He}{2} lines. In CO5.18, a strong \ion{C}{3} $\lambda5696$
emission line and a relatively weak \ion{C}{4} $\lambda5805$ emission line are
seen.  In CO9.09, a strong \ion{C}{4} line is seen but the \ion{C}{3}
$\lambda5696$ line is absent.  The \ion{O}{4} and \ion{O}{5} emission lines are
also found. CO5.18 and CO9.09 may correspond to WC8-10 and WC4/WO4 stars
respectively. Normalized spectra of all the models with the fiducial mass-loss 
rate are presented in Appendix \ref{subsec:Appendix2}.  The CO3.93 and CO2.16 models of
\citetalias{Yoon2019} show very similar normalized spectra to those of CO5.18 and CO5.50, respectively.

The effect of emission lines on the optical brightness can be observed when we
compare the full SED flux and the continuum flux. The optical magnitudes of $M_{F555W}$
calculated with them are also summarized in the inner tables in Figure
\ref{fig:sed}. The HE models have a similar $M_{F555W}$ for full spectra and continuum. 
For the CO models, however, the optical brightness with full spectra are
significantly  brighter than those with continuum. 
This implies that
the effect of emission lines is more important for SN Ic progenitors than SN
Ib progenitors.  The degree of this brightening by emission lines varies
non-linearly with  the wind density and the effective temperature, as discussed
in Section \ref{subsec:mdotvinfeffect}.

Note also that the optical brightness ($M_{F450W}$, $M_{F555W}$, $M_{F606W}$ 
and $M_{F814W}$) of the CO models are similar or somewhat lower
than those of the HE models, although the bolometric luminosities of most CO
models are systematically higher than the HE models (Table \ref{tab:input}, 
\ref{tab:output} and
the right panel of Figure \ref{fig:LM}). This qualitatively confirms the
conclusions of \citet{Yoon2012}, who argued that SN Ic progenitors would be
fainter than SN Ib progenitors. However, \citet{Yoon2012} adopt the black body
approximation with the hydrostatic surface temperature (i.e., BB-core case) and
they significantly underestimate the optical brightness of SN Ic progenitors. 

In Figure \ref{fig:windeffect}, we find that our fiducial models are
systematically fainter than Galactic WR stars.  This is mainly because our
models have a lower mass and bolometric luminosity than the Galactic WR stars,
on average.  However, the lowest mass HE models
(HE2.91 and HE2.97)  have optical magnitudes comparable to WR stars of $M \simeq 20
~M_\odot$ (i.e., Smith $v$ filter magnitude $M_{v} \simeq -5.2$), despite their
low bolometric luminosities (see Figure \ref{fig:LM}). This is because these
models have a large radius ($R_\star\simeq 20 - 25~R_\odot$) and a low surface
temperature ($T_\star \simeq 17000 - 19000~\mathrm{K}$), compared to those of
Galactic WR stars. These models are also brightest in the optical among our
models, implying that relatively low-mass helium star progenitors ($M \lesssim 3.0~
M_\odot$) would be relatively easy to detect in the visual wavelength range,
compared to more massive SN Ib/Ic progenitors \citep{Yoon2012, Kim2015, Folatelli2016, Eldridge2016, Yoon2017b}.

\subsection{Effects of the mass-loss rate and the wind terminal velocity} \label{subsec:mdotvinfeffect}

As discussed above in Section \ref{subsec:fiducial_models}, the effects of
winds on the optical brightness can be significant. However, the mass-loss rate
of SN Ib/Ic progenitors at the pre-SN stage is uncertain and might be different
from the mass-loss rate of the observed WR stars which would be mostly at core
helium burning stage \citep[cf.][]{Yoon2017b}.

\begin{figure*}[htp]
\includegraphics[width=0.8\textwidth]{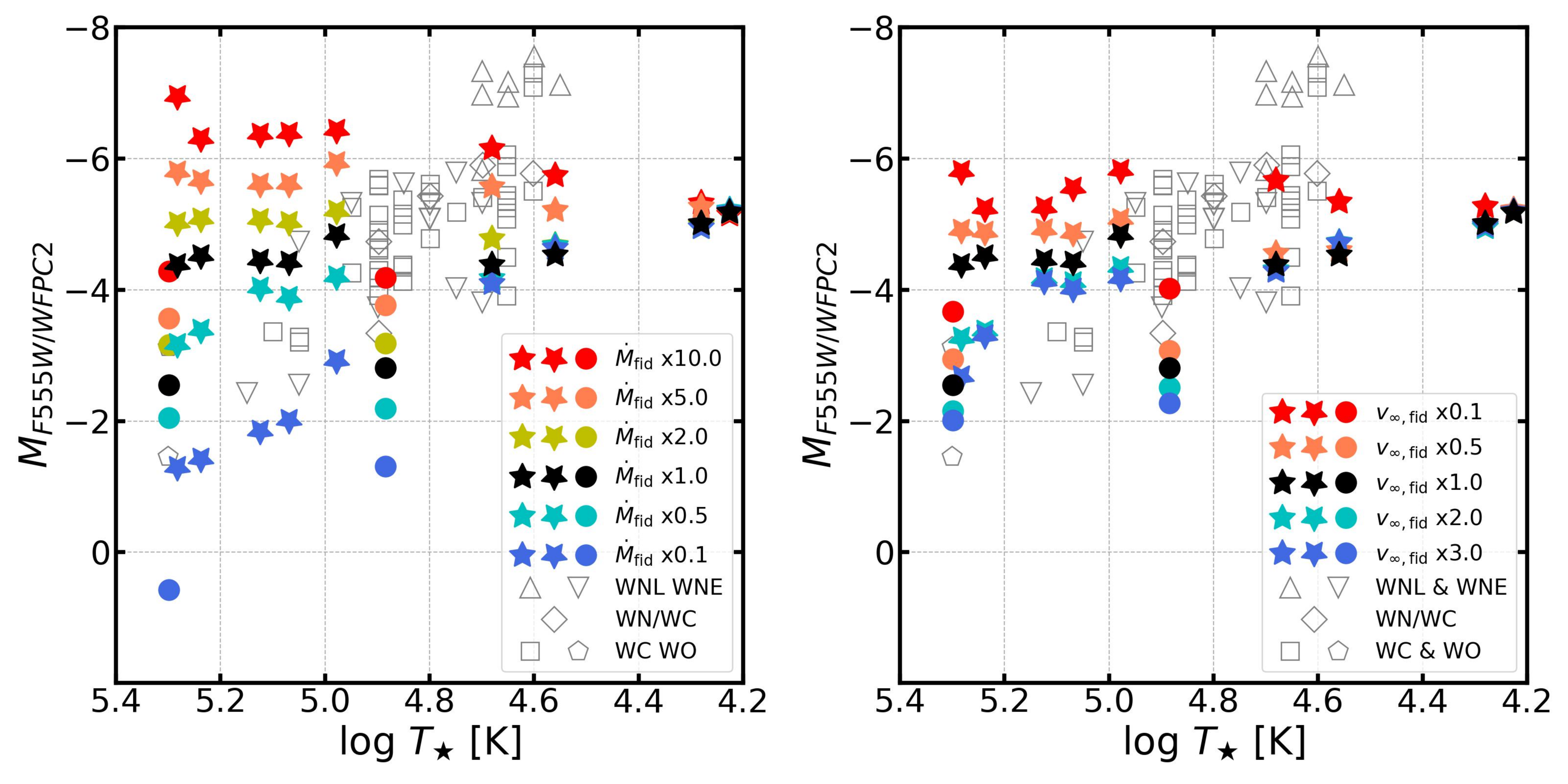}
\centering
\caption{$F555W$ ($HST$/WFPC2) filter magnitudes of SN Ib/Ic progenitors with various mass-loss rates (left) and wind terminal velocities (right). The upright and inverted stars denote the HE and CO progenitor models from \citetalias{Yoon2017b}, respectively. The CO progenitor models from \citetalias{Yoon2019} are presented with filled circles. Smith $v$ filter ($\lambda_\mathrm{peak}=5160\text{\AA}$) magnitudes of Galactic WR stars are plotted together for comparison \citep{Sander2019,Hamann2019}. Each color of the markers gives the adopted mass-loss rate and the wind terminal velocity as indicated by the labels.
\label{fig:mdot}}
\end{figure*}

We present the result of a parameter study with various mass-loss rates ($\dot{M} = 0.1~\dot{M}_\mathrm{fid} - 10~\dot{M}_\mathrm{fid}$) and wind terminal velocities ($v_\infty = 0.1~v_{\infty,\mathrm{fid}} - 3~v_{\infty,\mathrm{fid}}$)  in Figure \ref{fig:mdot}. Increasing the mass-loss rate or decreasing the wind terminal velocity implies a denser wind, which leads to a lower effective temperature and a higher brightness in the optical.  As seen in the figure, this effect is more prominent in the CO models (i.e., the models with $\log T_\star~\mathrm{[K]} > 4.8$ in the figure) than the HE models ($\log T_\star~\mathrm{[K]} < 4.8$ in the figure).  For example, an increase in $\dot{M}$ by a factor of 100 can lead to a 4 -- 6 mag difference in $M_{F555W}$ for CO models. This dependency of the optical brightness on the mass-loss rate allows us to find a constraint on the mass-loss rate of some SN Ib/Ic progenitors as discussed below (Section \ref{sec:implication}).  Normalized optical spectra of HE5.05 and CO5.18 with various mass-loss rates are presented in Appendix \ref{subsec:Appendix3}.  The figures show that optical properties and emission lines of CO5.18  are  more sensitively dependent on  the adopted mass-loss rate than the corresponding HE5.05 case. This is because the CO5.18 progenitor model is much more compact than HE5.05, making the effects of winds on the spectra more important.  

\begin{figure*}[htp]
\includegraphics[width=0.95\textwidth]{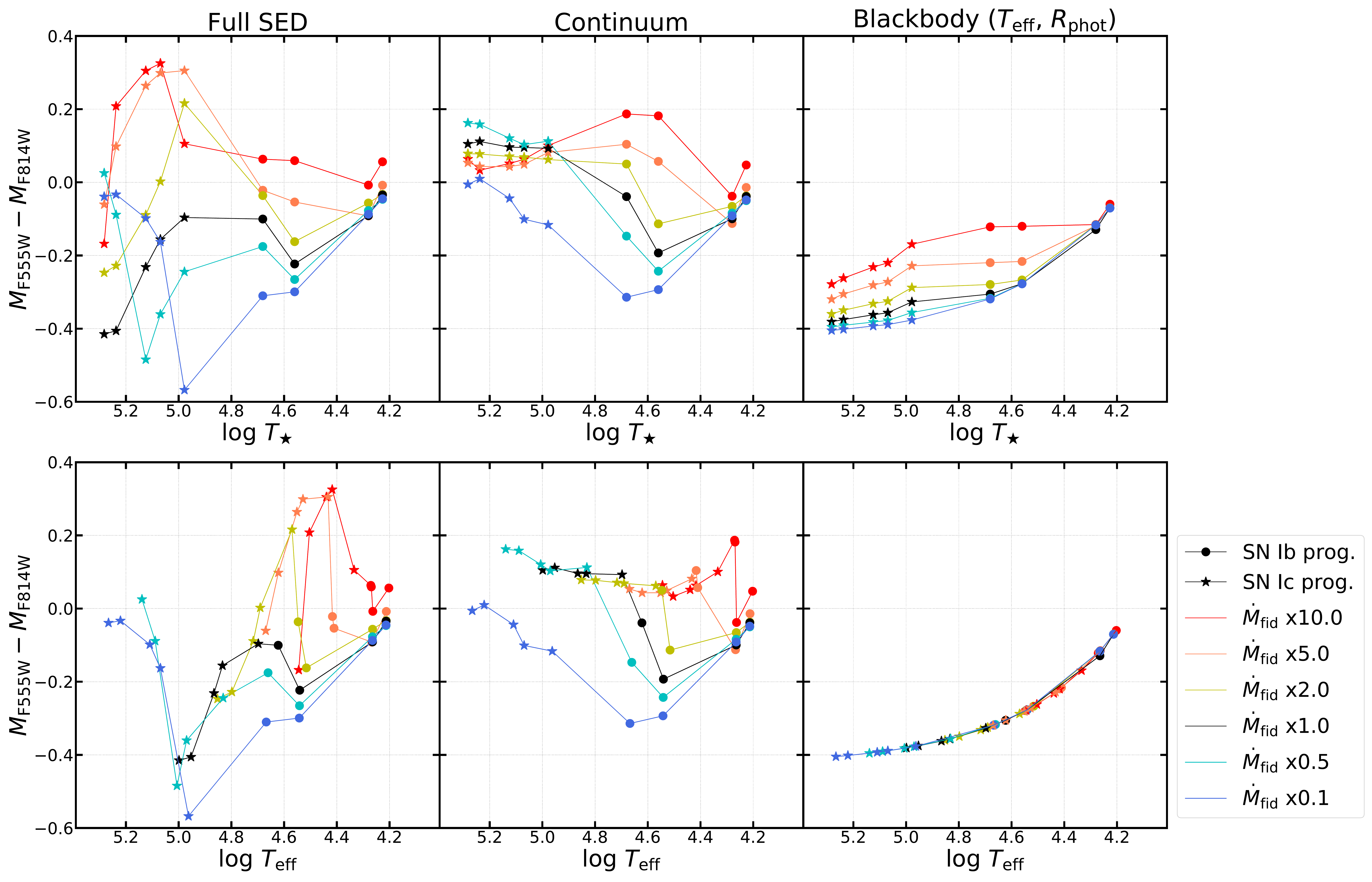}
\centering
\caption{Temperature-color relations for the SN Ib/Ic progenitor models from \citetalias{Yoon2017b}  (see Table \ref{tab:input}, \ref{tab:output} and \ref{tab:Appendix1}). The left panels show the $M_{F555W}-M_{F814W}$ ($HST$/WFPC2) colors calculated with full SEDs. The middle and right panels show the same colors calculated only with the continuum and black body fluxes at the photosphere, respectively. The circle and star markers denote SN Ib and Ic progenitor models, respectively. The color gives the adopted mass-loss rate as indicated by the labels. $\dot{M}_\mathrm{fid}$ means the fiducial mass-loss rate used in this paper. \label{fig:color}}
\end{figure*}

In Figure \ref{fig:color}, we present the color of the models for various wind mass-loss rates as functions of the hydrostatic surface temperature and the effective temperature.  The color would be a monotonic function of the effective temperature if the stars emitted black body radiation, as shown in the lower right panel of the figure. However, the continuum can be greatly affected by free-free emission as discussed above. The difference between middle and right panels in the figure shows the effect of free-free emission.  The continuum flux colors of  the CO models ($\log T_\star~\mathrm{[K]} > 4.8$ in the figure) are redder than the HE models ($\log T_\star~\mathrm{[K]} < 4.8$ in the figure) for the fiducial mass-loss rate, despite their higher $T_\mathrm{eff}$s.  For a given $T_\mathrm{eff}$, the continuum color is redder for a higher mass-loss rate due to the higher wind density and stronger free-free emission. The same result can also be found in \citet{Kim2015}.

Moreover, emission lines from the wind matter can also play an important role in determining the optical brightness in a specific filter. The difference between left and middle panels of Figure \ref{fig:color} shows the effect of emission lines. In many of our models with the fiducial mass-loss rate, more strong emission lines are found in the spectral range of the $F555W$ filter than in the $F814W$ filter range (see Figure \ref{fig:normalizedSED} and figures in Appendix), and it generally makes the color of the fiducial models with full SED bluer than the continuum color. The  CO5.50, CO6.17 and CO7.50 models  with a very high mass-loss rate (i.e., 5 and 10 times the fiducial value),  however, have a redder color with the full spectrum compared to the continuum color.  This is because of strong \ion{C}{2}/\ion{C}{3} emission lines in the $F814W$ filter wavelength range in these models. 

In short, the optical magnitude variation of SN Ib/Ic progenitors according to the wind density can be up to $\sim6$ mag for our considered parameter space (see Figure~\ref{fig:mdot}). The corresponding color change according to the effective temperature and the wind density is  highly non-monotonic and the progenitor color in terms of $M_{F555W}-M_{F814W}$  can be significantly redder (up to $\sim0.7$ mag) or bluer (up to $\sim0.2$ mag) depending on the progenitor structure and mass-loss rate, compared to the corresponding prediction under the black body approximation (see Figure \ref{fig:color}). This result indicates that prediction of the optical brightness and color of SN Ib/Ic progenitors using the black body approximation can lead to a significant error, and the effects of winds need to be properly considered to infer/constrain the progenitor properties from observations.

\subsection{Comparison with the Analytic Prediction} \label{subsec:theorycomparison}

\citet{deLoore1982} present a simple analytic prediction
for the photospheric radius and the effective temperature of a WR star having an optically thick wind. 
For example, assuming a constant opacity $\kappa$, $\beta = 2$  and $v(r = R_\star) = 0$ in the standard $\beta$-low wind velocity profile, 
the photospheric radius can be given as follows \citep{Langer1989b}:
\begin{fleqn}[\parindent]
\begin{equation}\label{eq:Reff_beta2}
    \begin{split}
        &R_\mathrm{phot}=R_\star+\frac{3\kappa|\dot{M}|}{8\pi v_\infty}~. 
    \end{split}
\end{equation}
\end{fleqn}
For $\beta=1$ as assumed in this study, we get
\begin{fleqn}[\parindent]
\begin{equation}\label{eq:Reff_beta1}
    \begin{split}
        &R_\mathrm{phot}={R_\star}\left[1-\mathrm{exp}\left(\frac{8\pi R_\star v_\infty}{3\kappa |\dot{M}|}\right)\right]^\mathrm{-1}~. 
    \end{split}
\end{equation}
\end{fleqn}
With this $R_\mathrm{phot}$,  
we can compute $T_\mathrm{eff}$ from the Stefan-Boltzmann
law (i.e., $T_\mathrm{eff}=T_\star(R_\star/R_\mathrm{phot})^\mathrm{1/2}$).

\begin{figure*}[htp]
\includegraphics[width=\textwidth]{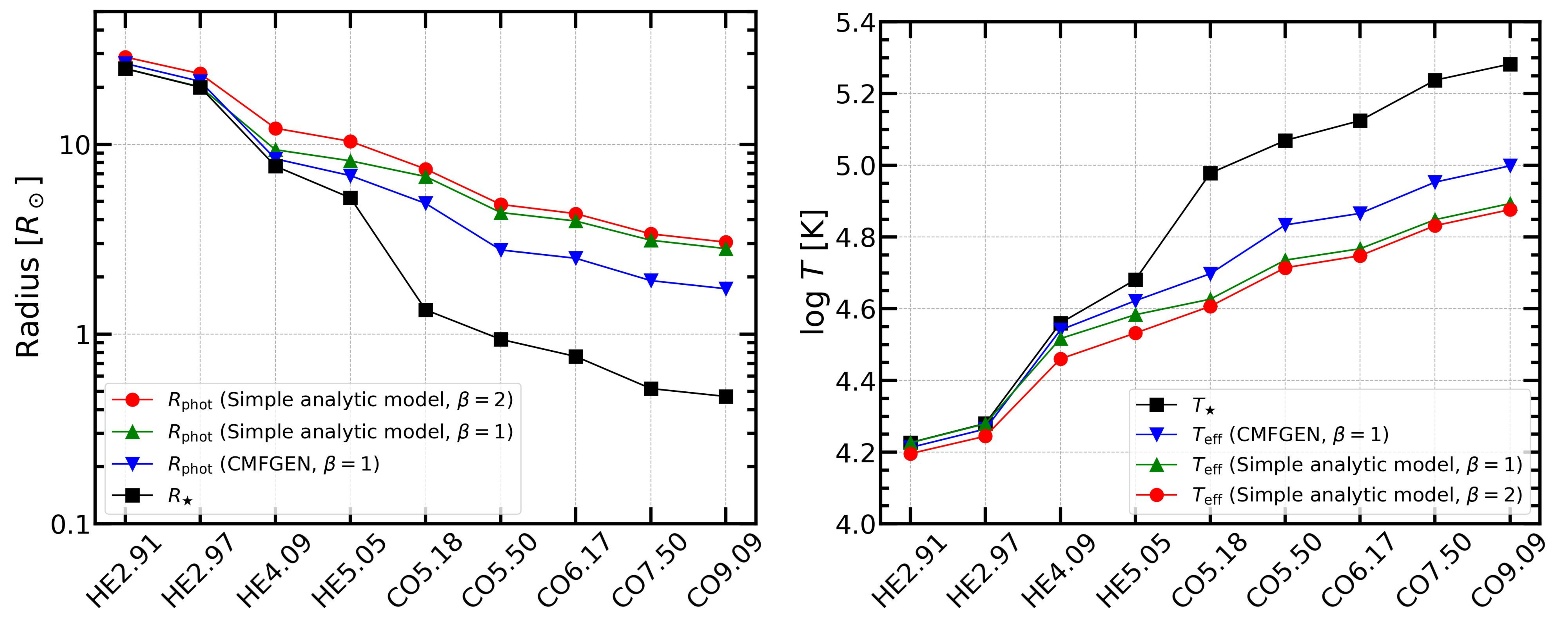}
\centering
\caption{Comparison of photospheric parameters of the SN Ib/Ic progenitor models calculated by the simple analytic model and the non-LTE CMFGEN code. The green triangle ($\beta=1$) and red circle ($\beta=2$) give the photospheric radius and the effective temperature calculated by the simple analytic model. The CMFGEN results are plotted with blue inverted triangles. The radius and temperature at the hydrostatic surface are denoted by black squares for comparison. \label{fig:theorycomparison}}
\end{figure*}

In Figure \ref{fig:theorycomparison}, we compare these analytically obtained
$R_\mathrm{phot}$ and $T_\mathrm{eff}$ for $\beta=1$ with the CMFGEN results.
We find that the two cases relatively well 
agree within 20 per cent, implying that the analytical prescription for 
$R_\mathrm{phot}$ and $T_\mathrm{eff}$ would be useful for a rough guess, 
particularly for SN Ib progenitors.  The caveat is that any prediction on
optical magnitudes and color from this $T_\mathrm{eff}$ under the black body
assumption could result in a significant error, as discussed above in
Sections \ref{subsec:fiducial_models} and \ref{subsec:mdotvinfeffect}.

\section{Implications for Supernova Progenitors} \label{sec:implication}

\subsection{Comparison with the detection limits of pre-explosion images} \label{subsec:detectionlimit}

\begin{figure*}[htp]
\centering
\includegraphics[width=0.95\textwidth]{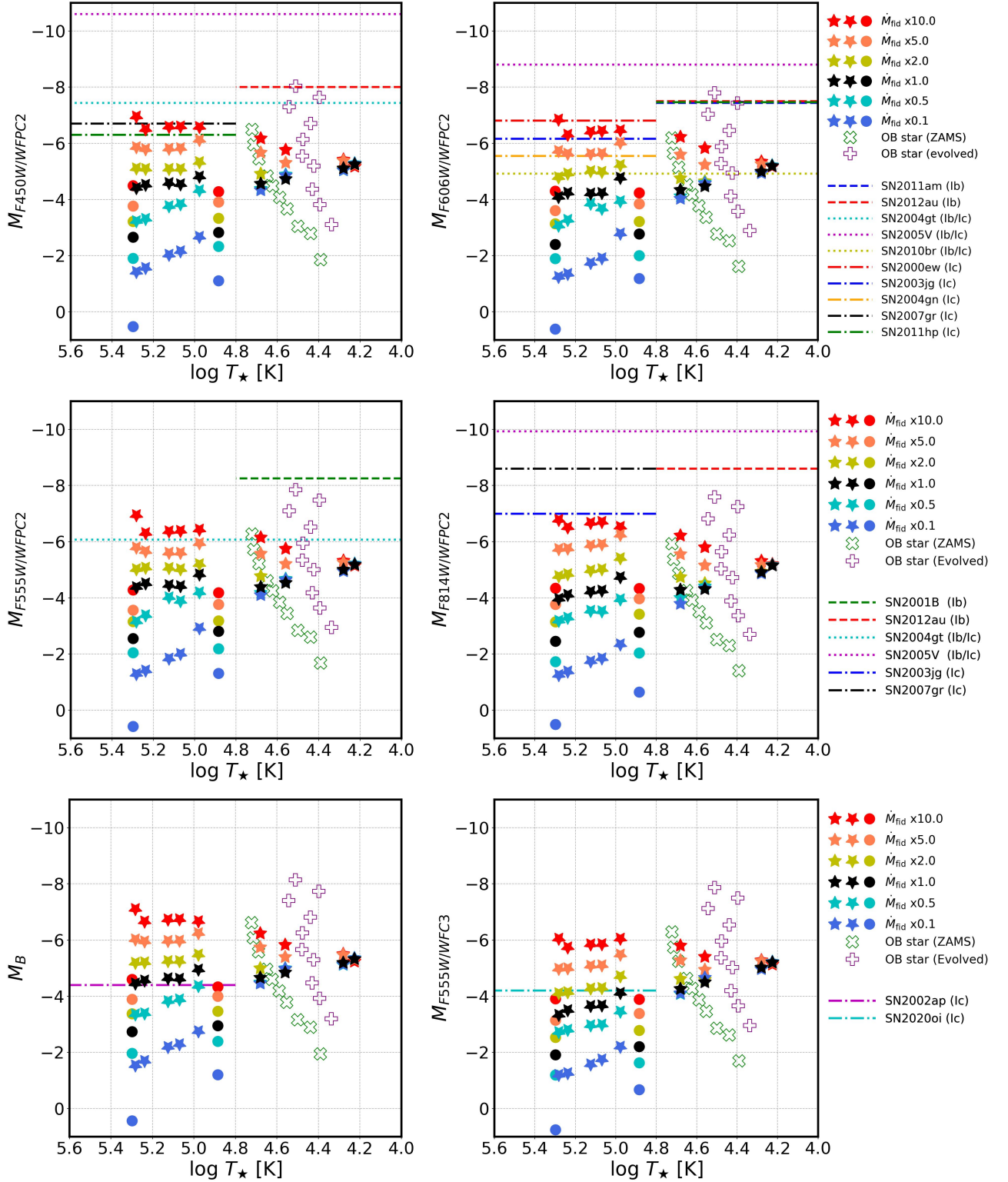}
\caption{Optical magnitude of progenitor models and the detection limits from the pre-explosion images of SNe Ib/Ic. Each panel shows the absolute magnitude of progenitor models in different filters : $F450W$, $F555W$, $F606W$ and $F814W$ in $HST$/WFPC2, Johnson-B and $F555W$ in $HST$/WFC3. The detection limits from the pre-explosion images of SNe Ib, Ic and Ib/Ic are plotted in dashed, dash-dotted and dotted lines, respectively \citep{Eldridge2013,Gagliano2021}. Optical magnitude of O/B type stars with 9, 12, 15, 20, 25, 32, 40, 60, 85 and 120 $M_\odot$ are plotted together in green (ZAMS) and purple (evolved) marker for the comparison \citep{Fierro2015}. Each color gives the adopted mass-loss rate as indicated by the labels.
\label{fig:obs}} 
\end{figure*}

There are only a limited number of candidates for SN Ib/Ic progenitors: SN Ib
progenitor candidates for iPTF13bvn \citep{Cao2013} and SN 2019yvr
\citep{Kilpatrick2021}, and  SN Ic progenitor candidiate for SN 2017ein
\citep{Vandyk2018, Kilpatrick2018}. In the cases of no direct identification, we can get the
detection limits from the pre-explosion images.  \citet{Eldridge2013}
report the detection limits for 12 SN Ib/Ic progenitors from the observations 
(See also \citet{Smartt2015} and \citet{Vandyk2017}). We add a detection limit of 
the pre-explosion image of SN Ic 2020oi \citep{Gagliano2021}. The detection limit is -4.2 mag
in $F555W$ filter ($HST$/WFC3) and obtained with the $3\sigma$ limit of the pre-explosion image.
In Figure \ref{fig:obs}, 
we present the detection limits in various filters. 
We compare the absolute magnitudes of our SN Ib/Ic progenitor models 
calculated with the same filter system. Note that the host galaxy extinction 
has not been considered in some of this sample~\citep{Eldridge2013, Gagliano2021}, 
in which case the detection limits might be lower than the given values. 
Even for the case where the host galaxy extinction was inferred, 
its uncertainty can be significant, which might also affect the estimated detection limits. 

In principle, we can roughly infer the upper limit of the mass-loss rate of
these non-detected progenitors because the optical brightness becomes
systematically higher for a higher mass-loss rate, in particular for SN Ic
progenitors.  We find that the observational detection limits except for
SN 2002ap are located above the CO2.16/CO3.93 model predictions (circle symbols) with  
$\dot{M}_\mathrm{fid}\times10.0$ in Figure \ref{fig:obs}. 
When compared to the relatively massive CO model predictions ($M>5.0~M_\odot$, 
inverted star symbols), the observational detection limits except
for 3 cases (SN 2002ap, SN 2010br and SN 2020oi) are located above the model
predictions with $\dot{M}_\mathrm{fid}\times2.0$. For SN 2010br
and the SN 2020oi, the mass-loss rate of the progenitor would not exceed twice
the fiducial mass-loss rate if their final masses were relatively high ($M\gtrsim5
~M_\odot$), while the upper limit of the mass-loss rate of the progenitor
would be higher than $M_\mathrm{fid}\times10$ if their final masses were relatively low ($M\lesssim4
~M_\odot$). The detection limit of SN 2002ap is -4.4 mag in Johnson-B filter,
which is larger than the fiducial model predictions of relatively massive CO progenitors ($M\gtrsim
5~M_\odot$) and smaller than the CO2.16/CO3.93 model cases. This implies
either that the SN 2002ap progenitor did not have a strong wind or that the
progenitor final mass was relatively low ($M \lesssim 4~M_\odot$). The
observationally inferred ejecta mass of SN 2002ap is about 2.5 $M_\odot$ \citep{Mazzali2007}, 
which is consistent with the latter interpretation.

There is another no direct identification case of Ca-rich SN Ib 2019ehk \citep{JacobsonGalan2020},
which is not presented in Figure \ref{fig:obs} and Table \ref{tab:detection_limit}. 
Its absolute detection limit from the pre-explosion image is -2.4 mag in $F555W$ filter ($HST$/WFPC2)
under the host galaxy distance assumption of d~$=16.2$ Mpc. The progenitor of SN 2019ehk is not detected 
although the detection limit is far higher than the HE model predictions ($M_{F555W}=-4\cdots-6$). 
This result is consistent with the scenario that Ca-rich SNe Ib would have a white dwarf origin
instead of massive stars~\citep[e.g.,][]{Perets2011,Kasliwal2012}.

\begin{deluxetable*}{lllll}
\tablenum{3}
\tablecaption{Upper limits of the mass-loss rate of the non-detected SN Ib/Ic progenitors in previous searches and constraints on the possible companion stellar mass\label{tab:detection_limit}}
\tablehead{
\colhead{Type} & \colhead{Name} & \colhead{$\dot{M}_\mathrm{max}$$^\mathrm{a}$} & \colhead{$M_\mathrm{comp,max}$$^\mathrm{b}$} & \colhead{$M_\mathrm{comp,max}$$^\mathrm{b}$} \\
\colhead{} & \colhead{} & \colhead{} & \colhead{(ZAMS)} & \colhead{(Evolved)} 
}
\startdata
\multirow{12}{*}{} 
SN Ib    & SN 2001B  &                                    &  &  \\ 
         & SN 2011am &                                    &  & $<85~M_\odot$  \\
         & SN 2012au &                                    &  & $<85~M_\odot$ \\
\hline
SN Ib/Ic & SN 2004gt &                                    & $<120~M_\odot$ & $<40~M_\odot$  \\
         & SN 2005V  &                                    &  & \\
         & SN 2010br &  $<\dot{M}_\mathrm{fid}\times5.0$  & $<60~M_\odot$  & $<20~M_\odot$ \\
\hline
SN Ic    & SN 2000ew &                                    &  & $<60~M_\odot$ \\
         & SN 2002ap &  $<\dot{M}_\mathrm{fid}\times1.0$  &  $<32~M_\odot$ & $<15~M_\odot$ \\
         & SN 2003jg &  $<\dot{M}_\mathrm{fid}\times10.0$ & $<120~M_\odot$ & $<40~M_\odot$ \\
         & SN 2004gn &  $<\dot{M}_\mathrm{fid}\times5.0$  & $<80~M_\odot$  & $<32~M_\odot$ \\
         & SN 2007gr &                                    &                & $<40~M_\odot$ \\
         & SN 2011hp &  $<\dot{M}_\mathrm{fid}\times10.0$ & $<120~M_\odot$ & $<40~M_\odot$ \\
         & SN 2020oi &  $<\dot{M}_\mathrm{fid}\times2.0$  & $<32~M_\odot$  & $<15~M_\odot$ \\
\enddata
\tablecomments{Detection limits are from the $HST$/WFPC2 images except SN 2002ap and SN 2020oi from Johnson-B  and $HST$/WFC3 image, respectively. \newline
$^\mathrm{a}$Upper limits of the mass-loss rates are obtained by comparing with HE and relatively massive ($M\gtrsim5~M_\odot$) CO model predictions. Blank if a detection limit is brighter than the progenitor model with $\dot{M}_\mathrm{fid}\times10.0$.\newline
$^\mathrm{b}$Blank if a detection limit is brighter than the ZAMS or evolved $120M_\odot$ O type star.}
\end{deluxetable*}

We obtain the upper limit of the progenitor mass-loss rate
($\dot{M}_\mathrm{max}$) by comparing various mass-loss rate models with 
detection limits of pre-explosion images, and summarize the results in Table
\ref{tab:detection_limit}. These upper limits are obtained by comparing 
the detection limits with HE and relatively massive ($M\gtrsim5~M_\odot$) CO model predictions. 
All of our relatively low-mass CO2.16/CO3.83 models are below the detection limits and cannot provide a upper limit for our considered parameter space.
This result implies that if the progenitors of observed SNe Ib/Ic had a mass-loss rate comparable
to $\dot{M}_\mathrm{fid}$, non-detection would be a natural consequence for most cases, and
that deeper observations with an optical absolute magnitude larger than about $-4
\cdots -5$ are needed to directly observe the progenitors and to make a better
constraint on their properties. The point source magnitude limit with 
an optical wide band filter in $HST$/WFC3 for 1 hour observation is $\sim27.1-27.9$ mag 
(at a signal-to-noise ratio of 10). This means that a progenitor with the absolute 
optical magnitude of $-4$ 
mag would be detected only if it were in a galaxy closer than $\sim16-24$ Mpc. 
The limiting distance can be decreased to $\sim10-15$ Mpc for 10 min observation.  
The host galaxy extinction should be additionally considered for a more precise distance limit.

Many SN Ib/Ic progenitors would be produced in binary systems and the
contribution of the companion star to the optical brightness would also be
significant. We take CMFGEN model spectra of O/B type stars at zero-age main
sequence (ZAMS) and an evolved phase on the main sequence having the initial
mass range of $9-120~M_\odot$ from \citet{Fierro2015}, and present their optical
magnitudes in Figure \ref{fig:obs}.  Our fiducial progenitor models have
optical magnitudes comparable to those of $15-25~M_\odot$ evolved O stars or
$25-60~M_\odot$ O stars at ZAMS.  For each progenitor, we get the upper limit of
the possible companion star mass ($M_\mathrm{comp,max}$) by comparing the
detection limit with the O/B star magnitudes, ignoring the contribution from
the SN progenitor. The result is summarized in Table \ref{tab:detection_limit}.
Even for the deepest detection limit cases, we can
only exclude a companion star more massive than $15~M_\odot$. 

\citet{Zapartas2017} predict the possible companions of stripped-envelope SNe
at the explosion time using the population synthesis simulation. The result of
the study shows that only $\sim$ 20~\% of stripped-envelope SNe progenitors
would have massive a companion with $>20~M_\odot$.  This result also implies
that most SN Ib/Ic progenitors would not have massive companion star that is
bright enough to exceed the previous detection limits.

\subsection{Comparison with the progenitor candidate of SN Ib iPTF13bvn} \label{subsec:iPTF13bvn}

The progenitor candidate of Type Ib SN iPTF13bvn in NGC 5806 is reported by \citet{Cao2013} and later convincingly identified as the progenitor by a follow-up observation by \citet{Eldridge2016, Folatelli2016}. \citet{Cao2013} measure the Milky Way and host galaxy reddening as $E(B-V)_{MW}=0.0278$ and $E(B-V)_{host}=0.0437$, respectively, from the \ion{Na}{1} D lines and use the adopted distance modulus of $\mu=31.76\pm0.36~(22.5\pm2.4~$Mpc)
\citep{Tully2009}. Their photometry of the progenitor candidate gives the values of $m_{F435W}=26.50~\pm~0.15$, $m_{F555W}=26.40~\pm~0.15$ and $m_{F814W}=26.10~\pm~0.20$ mag in the $HST$ Advanced Camera for Surveys (ACS) filter system.

On the other hand, \citet{Bersten2014} suggest higher extinction values of $E(B-V)_{MW}=0.0447$ and $E(B-V)_{host}=0.17$ using the SN light curve and \ion{Na}{1} lines. Moreover, \citet{Eldridge2015} give brighter photometry values of $m_{F435W}=25.80\pm0.12$, $m_{F555W}=25.80\pm0.11$ and $m_{F814W}=25.88\pm0.24$ mag for the same progenitor candidate. In this study, we present the results of every combination of different extinctions and photometry values.

\begin{figure*}[t]
\includegraphics[width=\textwidth]{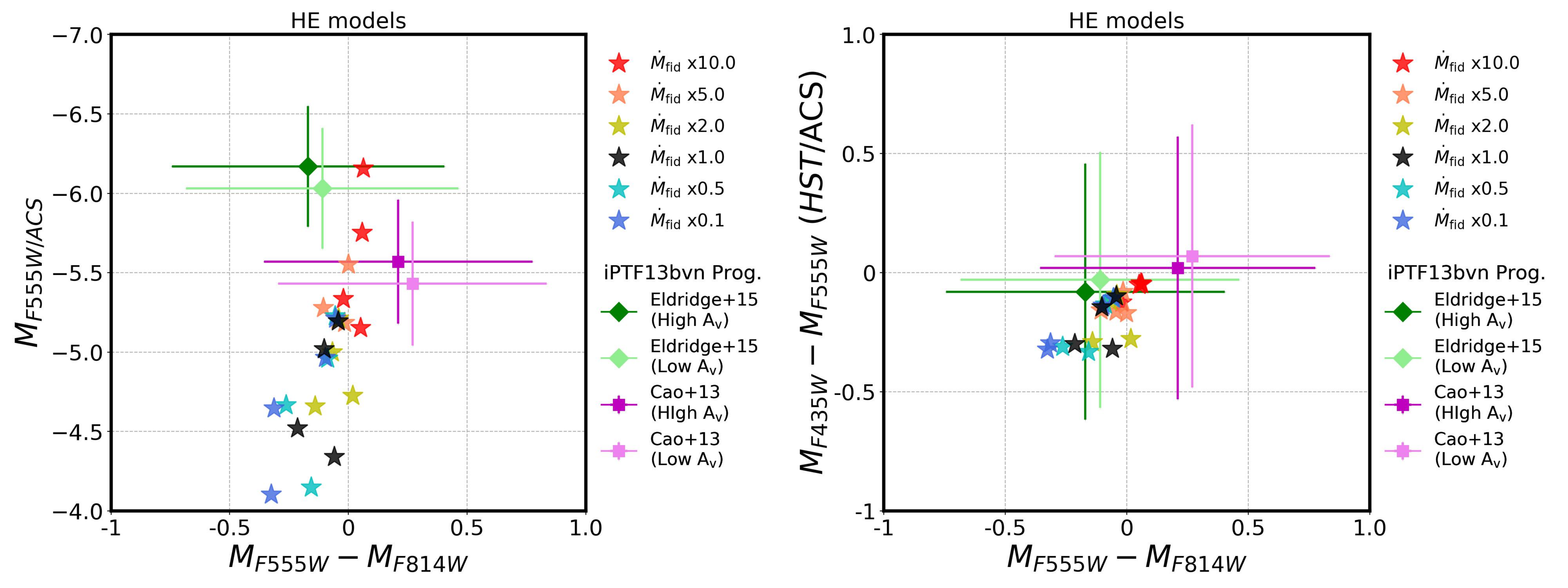}
\caption{Color-magnitude diagram (left) and color-color diagram (right) of the HE models. The absolute optical magnitudes and colors of the iPTF13bvn progenitor candidate for two different extinction and photometry values are presented together. Magenta and green markers denote the values calculated with the photometry results of \citet{Cao2013} and \citet{Eldridge2015}, respectively. The values with high and low extinction assumptions are plotted in dark and light colors, respectively.
\label{fig:Ib_CMD}} 
\end{figure*}

In Figure \ref{fig:Ib_CMD}, we show our HE progenitor models and the iPTF13bvn
progenitor candidate in the color-magnitude diagram and the color-color
diagram. We present four observational results together. For the photometry 
result of \citet{Cao2013}, the magnitude and the color of iPTF13bvn progenitor can be
explained by the relatively low-mass fiducial models HE2.91 and HE2.97
or by most HE models with $\dot{M}_\mathrm{fid}\times5\cdots10$.  
For the photometry result of \citet{Eldridge2015}, however, the progenitor candidate of the
iPTF13bvn is brighter than most of our considered HE models, being about 1 mag
brighter than the fiducial models in the optical. Only the
HE4.09 and HE5.05 models with $\dot{M}_\mathrm{fid}\times10.0$ can explain the optical brightness.
The inferred ejecta mass ($\sim2.0-2.3~M_\odot$) and helium core mass ($\sim3.4-3.5~M_\odot$) 
of iPTF13bvn \citep{Bersten2014, Fremling2014} are in range of our HE models.

\begin{figure*}[t]
\includegraphics[width=\textwidth]{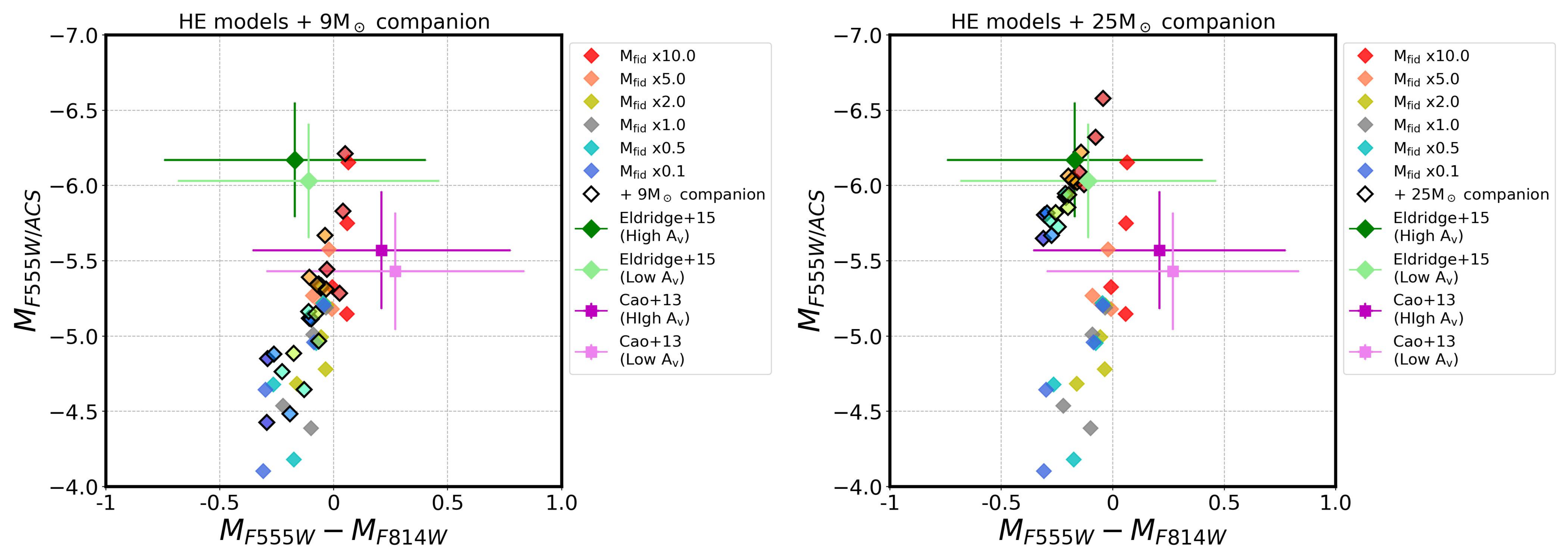}
\caption{Color-magnitude diagrams of the HE progenitor models. The left and right panel give the composite results with a 9$M_\odot$ and a 25$M_\odot$ companion star, respectively. The diamond markers without a black box denote the progenitor models without a companion star, and the filled diamonds with black border  denote the composite models with a companion star. Each color of the markers gives the adopted mass-loss rate as indicated by the labels. The values of the iPTF13bvn progenitor candidate are plotted together as in Figure \ref{fig:Ib_CMD}.
\label{fig:Ib_companion}} 
\end{figure*}

To consider the possible contribution from a companion star of the progenitor,
we also calculate the composite magnitudes from the spectral models of both the
HE progenitors and the evolved main sequence stars of $M_\mathrm{init}=9$ and
$25~M_\odot$ from \citet{Fierro2015}. As seen in Figure \ref{fig:Ib_companion},
the optical properties of the iPTF13bvn progenitor candidate are consistent
with the models with a $9~M_\odot$ or $25~M_\odot$ companion star for photometry
result of \citet{Cao2013} and \citet{Eldridge2015}, respectively. A $9~M_\odot$
companion would be a reasonable solution within the standard binary scenario of
SN Ib progenitors \citep[e.g.,][]{Bersten2014, Yoon2017b} and agrees well with
other previous studies on iPTF13bvn \citep[e.g.,][]{Fremling2014,
Srivastav2014, Kuncarayakti2015, Fremling2016, Folatelli2016, Eldridge2016}.  A
companion star as massive as $25~M_\odot$ is somewhat hard to explain. Given
that the inferred ejecta mass of iPTF13bvn is only about $2.0~M_\odot$
, such a massive companion in a binary system
could be produced only with conservative mass transfer and an unusually high
mass-loss rate from the naked primary star during the post-mass transfer phase
\citep{Wellstein1999}.  On the other hand, \citet{Groh2013a} suggest a single
WR star progenitor (initial mass $M_\mathrm{init}=31-35~M_\odot$) by comparing
of the Geneva single star evolution model with the observation., but the
predicted pre-explosion mass of $M\simeq10.9~M_\odot$) is incompatible with the
inferred ejecta mass of iPTF13bvn. 

\subsection{Comparison with the progenitor candidate of SN Ib 2019yvr} \label{subsec:SN2019yvr}

\citet{Kilpatrick2021} report the SN Ib progenitor candidate of SN 2019yvr,
which is found $\sim980$ to 870 days ($\sim 2.6$ years) before the explosion in a $HST$/WFC3 observation.
The average photometry across all epochs in four bands gives the results of
$m_{F438W}=26.138~\pm~0.162$, $m_{F555W}=25.351~\pm~0.032$,
$m_{F625W}=24.897~\pm~0.022$ and $m_{F814W}=24.253~\pm~0.032$ mag. The color
corrected only with the Milky Way extinction $E(B-V)_{MW}=0.02$ is
$m_{F555W}-m_{F814W}=1.065~\pm~0.045$ mag, which corresponds to the
temperature of $T_\mathrm{eff}=3340$ K. Without a fairly high host galaxy
extinction, the color is too red to be explained by our HE progenitor models (see
the values in Figure \ref{fig:Ib_CMD}). 

\begin{figure*}[ht]
\includegraphics[width=0.8\textwidth]{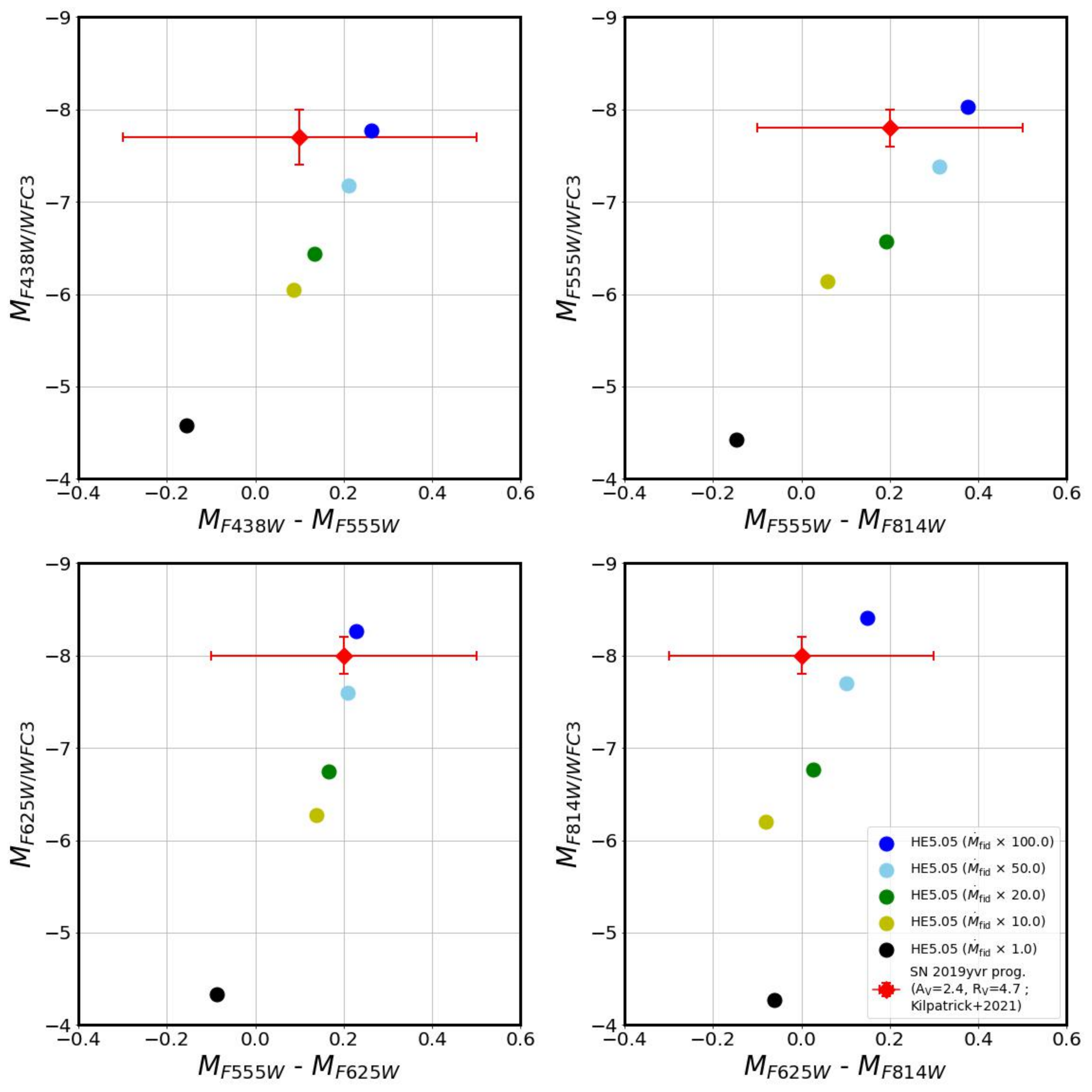}
\centering
\caption{Color-magnitude diagrams of SN 2019yvr progenitor candidate comparing with SN Ib progenitor models. SN 2019yvr progenitor candidate and HE5.05 models are plotted in diamond and circles, respectively. The color denotes the adopted mass-loss rate of HE5.05 models as indicated by the labels. Color and absolute magnitude of SN 2019yvr progenitor are corrected under the assumption of $A_\mathrm{v}=2.4$ and $R_\mathrm{v}=4.7$. \label{fig:2019yvr}} 
\end{figure*}

\citet{Kilpatrick2021} obtained the high host galaxy extinction value of
$A_V=2.4^{+0.7}_{-1.1}$ and $R_V=4.7^{+1.3}_{-3.0}$ from \ion{Na}{1} D spectra
and color curves of SN 2019yvr. They infer the physical parameters of the
progenitor candidate by fitting with the single star SEDs from
\citet{Pickles2010}. The best-fitting values are $T_\mathrm{eff}=6800$ K,
log($L/L_\odot$)$=5.3~\pm~0.2$ and $R_\mathrm{phot}=320~R_\odot$ from F2I star
model, and they argue that this large radius of the photosphere might be
explained by a very high mass-loss rate of $\sim10^{-4}~M_\odot$ $\mathrm{yr^{-1}}$. To
check this possibility, we construct spectra of our most massive HE model
(HE5.05) with mass-loss rates up to $100 \times \dot{M}_\mathrm{fid}$ and
present the results in Table \ref{tab:2019yvr_tab} and
Figure~\ref{fig:2019yvr}. We find that the models with very high mass-loss
rates (i.e., $\dot{M}  = 50\cdots 100\times \dot{M}_\mathrm{fid}$) can explain
the optical color and magnitudes of the progenitor candidate. 
The SN 2019yvr progenitor 
candidate is observed during 980 to 870 days before the explosion.   
Although stellar evolution models predict that the surface properties of SN Ib progenitors
would not undergo a rapid change during the pre-explosion phase 
\citep{Yoon2017a}, 
the inferred high mass-loss rate implies that the progenitor underwent
mass eruption shortly before the explosion, for which energy injection from the inner covective layers via wave heating 
during core neon/oxygen burning might be a possible explanation
 \citep{Fuller2018, Leung2021}. Note also 
that evidence for pre-SN mass eruption is found in the early-time light curves 
of some stripped-envelope supernovae~\cite[e.g.,][]{Jin2021}. 
However, the observational signature of late time interaction of SN 2019yvr~\citep{Kilpatrick2021}
is not likely to be related to this mass eruption because the matter emitted during the last several years
 would be confined to a small radius ($\lesssim 10^{15}$ cm) from the progenitor at the time of explosion and
would affect only the early-time SN light curve~\citep[see][]{Jin2021}.

\begin{deluxetable*}{ccccccccc}[ht!]
\tablenum{4}
\tablecaption{Physical parameters of the progenitor candidate of SN 2019yvr and SN Ib progenitor models \label{tab:2019yvr_tab}}
\tablehead{
\colhead{} & \colhead{} & \colhead{SN 2019yvr} & \colhead{HE5.05} & \colhead{HE5.05} & \colhead{HE5.05} & \colhead{HE5.05} & \colhead{HE5.05}\\ 
\colhead{} & \colhead{} & \colhead{\citep{Kilpatrick2021}} & \colhead{($\dot{M}_{\mathrm{fid}}$)} & \colhead{($\dot{M}_{\mathrm{fid}}\times10$)} 
& \colhead{($\dot{M}_{\mathrm{fid}}\times20$)}   &  \colhead{($\dot{M}_{\mathrm{fid}}\times50$)} &  \colhead{($\dot{M}_{\mathrm{fid}}\times100$)} 
}
\startdata
\multirow{4}{*}{} 
                      log $L/L_\odot$ &     & 5.3 & 5.1 & 5.1 & 5.1 & 5.1 & 5.1 \\
                      $\dot{M}$ & [$M_\odot$ yr$\mathrm{^{-1}}$]          & $>$1.4e-04 & 8.13e-06 & 8.13e-05 & 1.63e-04 & 4.06e-04 & 8.13e-04\\
                      $T_{\mathrm{eff}}$ & [K]  & 6800 & 41910 & 18660 & 14260 & 10230 & 7852\\
                      $R_{\mathrm{phot}}$ & [$R_{\odot}$] & 320 & 6.834 & 34.49 & 59.03 & 114.7 & 194.7 \\
\enddata
\tablecomments{First column shows the physical parameters of the best-fit single star progenitor model (F2I model from \citet{Pickles2010}) for SN 2019yvr from \citet{Kilpatrick2021}. The other columns show the physical parameters of HE5.05 model with various mass-loss rates. HE5.05 model is evolved from a 8~$M_\odot$ helium star, which corresponds to a $\sim25~M_\odot$ star at ZAMS.
}
\end{deluxetable*}

Here we obtain the extinction corrected absolute magnitude of the progenitor
candidate with the host galaxy extinction ($A_V=2.4$ and $R_V=4.7$) and distance
modulus ($m-M=30.8~\pm~0.2$ mag) adopted from \citet{Kilpatrick2021}.  The
corrected values are $M_{F438W}=-7.7~\pm~0.3$, $M_{F555W}=-7.8~\pm~0.2$,
$M_{F625W}=-8.0~\pm~0.2$ and $M_{F814W}=-8.0~\pm~0.2$ mag for each \textit{HST}/WFC3
filter.  The adopted host galaxy extinction values, however, are obtained from
the observation after the SN explosion, so we have to consider the possibility
of additional extinction sources (e.g.  dusty circumstellar matter) before the
explosion. If there were additional extinction sources, the corrected
magnitudes and the color would move to the brighter and bluer direction away
from the model predictions.

\subsection{Comparison with the progenitor candidate of SN Ic 2017ein} \label{subsec:SN2017ein}

\begin{figure*}[thb!]
\includegraphics[width=\textwidth]{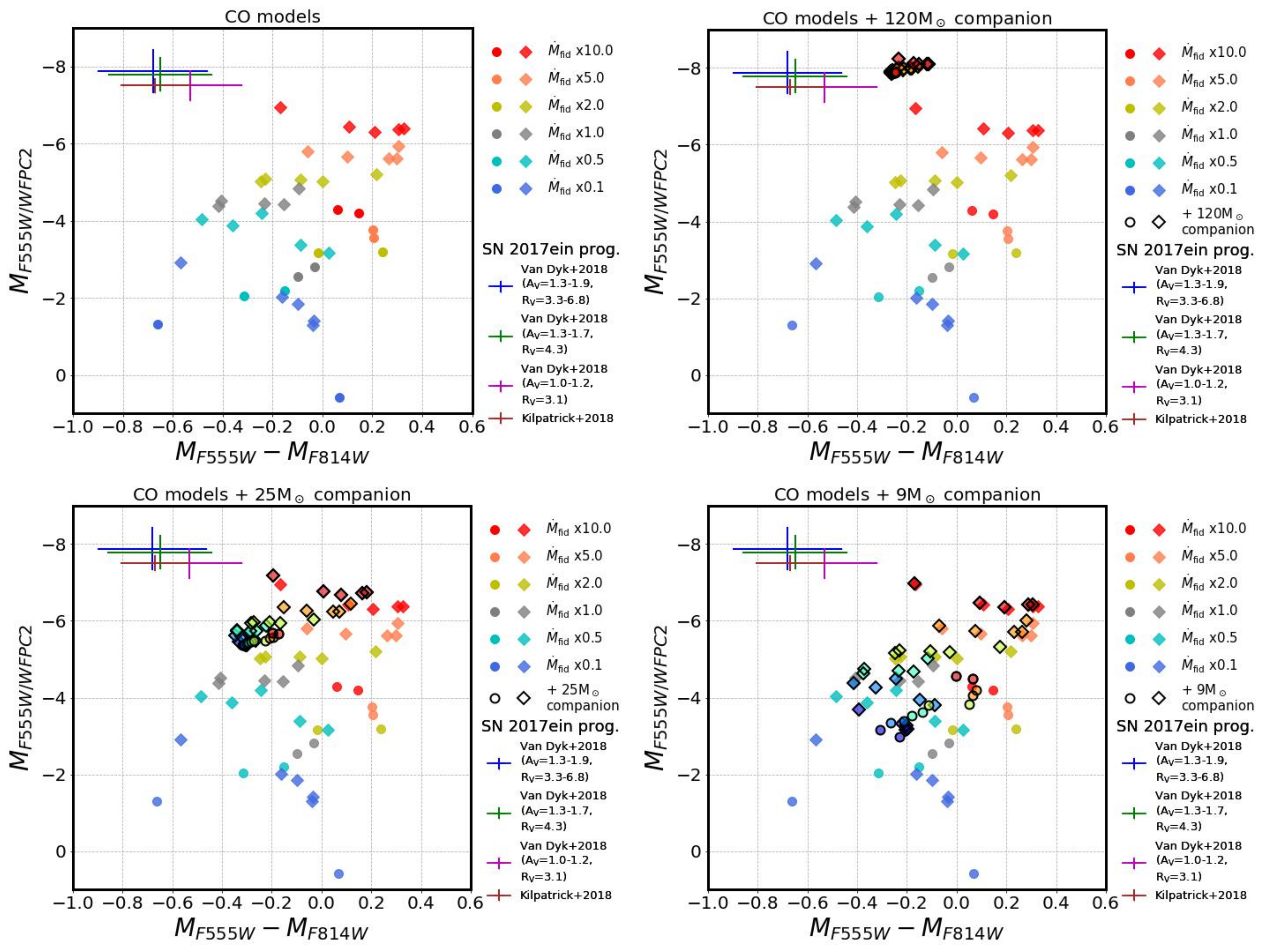}
\caption{Color-magnitude diagrams of CO progenitor models. The upper left panel gives the results without considering a companion star. The upper right, lower left, and lower right panels present the composite results with a 120, 25, and 9~$M_\odot$ companion star, respectively. The diamond and circle marker with/without a black border denotes the progenitor models with/without a companion star. The reported magnitude and color of SN 2017ein progenitor candidate are plotted together for comparison. Each color gives the adopted mass-loss rate as indicated by the labels.\label{fig:CMD_Ic}} 
\end{figure*}

There is a progenitor candidate for SN Ic 2017ein. \citet{Kilpatrick2018} suggest the Milky Way extinction of $A_V=0.058$ $(R_V=3.1)$ and the host galaxy extinction of $A_V=1.2$ $(R_V=2.6)$. With the distance modulus $\mu=31.17~\pm~0.10$ for NGC 3938 \citep{Tully2009}, the extinction corrected magnitudes are obtained as $M_{F555W}=-7.5~\pm~0.2$ mag and $M_{F814W}=-6.7~\pm~0.2$ mag for each $HST$/WFPC2 filter. 
\citet{Vandyk2018} independently analyze the $HST$ photometric data before the explosion, and obtain a similar result with various extinction assumptions, 
as shown in Figure \ref{fig:CMD_Ic}. In the figure, this candidate is compared with our CO progenitor models, the progenitor candidate is much more luminous than our fiducial models in the optical (by about 2 mag) and only a very high mass-loss rate of $\dot{M}_\mathrm{fid}\times10.0$ can result in a barely comparable magnitude in the $F814W$ filter. The observed color ($M_{F555W}-M_{F814W}$) is very blue compared to our progenitor models. No model within our parameter space satisfies both the observed magnitude and color of this candidate. 

We also consider O/B type evolved main-sequence companion stars of 9, 25, and
120~$M_\odot$ whose spectra are taken from \citet{Fierro2015} and the composite magnitudes and colors with the CO progenitor models are plotted in Figure \ref{fig:CMD_Ic}. Since our most massive CO progenitor
model (CO9.09) is evolved from a 25$M_\odot$ He star, which corresponds to
$M_\mathrm{init}\approx65~M_\odot$, a companion star with 120~$M_\odot$ is not realistic.
The upper right panel of Figure \ref{fig:CMD_Ic} shows, however, that even
this unrealistically massive companion cannot explain the blue color of the
progenitor candidate.

On the other hand, all of the composite models with 9 and 25~$M_\odot$
companion stars are fainter and redder than the candidate (the lower panels of
Figure \ref{fig:CMD_Ic}).  Under the assumption that the values inferred from
the observation are reliable, we conclude that the reported candidate of SN
2017ein is too blue and too bright in the optical to be explained by a typical
SN Ic progenitor.

\citet{Kilpatrick2018} suggest a 55~$M_\odot$ single star or a $80+48~M_\odot$ binary at ZAMS
as the progenitor. \citet{Vandyk2018} suggest a $\sim47-48~M_\odot$ single star or a $\sim60-80~M_\odot$ in binary at ZAMS. 
Our SN Ic progenitor models cover this initial 
mass range since our most massive model (CO9.09) corresponds to a $60-70~M_\odot$ star at ZAMS.
\citet{Kilpatrick2018} obtain a host galaxy extinction value of
$A_\mathrm{V}=1.2-1.9$ using \ion{Na}{1} lines. The extinction corrected color
and magnitude of the SN 2017ein progenitor candidate are calculated using the lowest value of
$A_\mathrm{V}=1.2$, but it is already too bright and blue. 
Moreover, there can be additional extinction sources before the explosion,
which can make it move to brighter and bluer region.
\citet{Vandyk2018} presents a similar result in spite of the various 
extinction assumptions.
To confirm if the observed candidate was the real progenitor or
not, we need additional observation to check the disappearance of the
candidate.

\section{Conclusions} \label{sec:conclusion}

We present stellar atmosphere models of SN Ib/Ic progenitors having different
chemical compositions  for the pre-SN mass range of $2.16 - 9.09~M_\odot$, which are
calculated by the non-LTE radiative transfer code CMFGEN.  We investigate the
effects of various wind parameters on the resulting spectra and optical
properties and discuss implications of our results for direct identification of
SN Ib/Ic progenitors. 

Our results indicate that optical properties of SN Ib/Ic progenitors can be
greatly affected by the presence of optically thick wind, and inferring SN
Ib/Ic progenitor masses and bolometric luminosities from optical brightness and
colors observed in pre-SN images  could be misleading.   More specifically, we
draw the following conclusions from our discussions. 

\begin{enumerate}

\item 
Presence of sufficiently high density wind material around a SN Ib/Ic
progenitor can make the photosphere significantly lifted-up from the
hydrostatic stellar surface.  Strong emission lines and free-free emission
from the optical thick wind  can also critically affect optical magnitudes. 
These effects would generally make SN Ib/Ic
progenitors  brighter in the optical compared to the corresponding case
without winds, in particular for helium-deficient compact SN Ic progenitors.
Our models predict that SN Ic progenitors would be brighter by more than 3 mag
in the optical  than the case without the wind effects~for the fiducial WR 
mass-loss rate.  On the other hand, optical
properties of SN Ib progenitors having an extended helium-rich envelope are not
greatly affected by the effects of winds for the fiducial WR mass-loss rate.
The optical brightness becomes higher for a higher mass-loss rate (or a lower
wind terminal velocity), and this tendency is found more clearly for a more
compact progenitor (Sections \ref{subsec:fiducial_models} and
\ref{subsec:mdotvinfeffect}). 

\item 
The wind effects also greatly  affect the color of SN Ib/Ic progenitors.  In
general, the color becomes redder for a higher mass-loss rate (or a lower wind
terminal velocity) because of the lifting-up of the photosphere and the
continuum excess due to free-free emission in a relatively long wavelength
range.  However, the optical colors are not well correlated with the effective
temperature, nor with the hydrostatic surface temperature. This is because strong
emission lines found in specific filters and the free-free emission make the
color-temperature relations highly non-monotonic (Sections
\ref{subsec:fiducial_models} and  \ref{subsec:mdotvinfeffect}). 

\item For a typical mass-loss rate of WR stars, the optical brightness of most
SN Ib/Ic progenitors would not be high enough to be detected within the
detection limits of previous searches in pre-SN images.  Also, these detection
limits were not deep enough to observe companion stars of
$M_\mathrm{comp}<15~M_{\odot}$, which would be the typical case for SN Ib/Ic
progenitors.  A deep search with an optical absolute magnitude larger than $\sim -4$
would be needed for identification of most of oridinary SN Ib/Ic progenitors.  
The distance limit for a direct progenitor detection is $\sim10-15$ Mpc
and $\sim16-24$ Mpc for 10 min and 1 hour observation, respectively, 
for an optical wide band filter of $HST$/WFC3.
(Section \ref{subsec:detectionlimit}). 

\item The optical brightness and color of the observed SN Ib iPTF13bvn
progenitor can be explained by our models within the considered parameter
space.  The photometry result of \citet{Cao2013} is consistent with  the
relatively low-mass fiducial HE models (HE2.91 and HE2.97), HE4.09/HE5.05
models with a relatively high mass-loss rate
($\dot{M}_\mathrm{fid}\times5\cdots10$) or existence of a $\sim 9~M_\odot$
companion star.  For the photometry result of \citet{Eldridge2015}, the optical
brightness and color of the progenitor can be explained by HE4.09/HE5.05 models
with a very high mass-loss rate ($\dot{M}_\mathrm{fid}\times10$) or with the
existence of $\sim 25~M_\odot$ companion star~(Section \ref{subsec:iPTF13bvn})

\item The SN Ib 2019yvr progenitor candidate is very bright and red in the optical
compared to our fiducial HE models. Its optical brightness and color  can be
explained by the HE5.05 model with an unusually high mass-loss rate of
$\dot{M}_\mathrm{fid}\times50\cdots100$ This implies the SN 2019yvr progenitor
experienced mass-loss enhancement shortly before the SN explosion  (Section
\ref{subsec:SN2019yvr}). 

\item The SN Ic 2017ein progenitor candidate is too blue and bright to be
explained with any of our models whether we consider a companion star or not.
Any known massive stars could not explain both the optical brightness and the
blue color of this progenitor candidate (Section \ref{subsec:SN2017ein}).

\begin{acknowledgements}
This work has been supported by the National Research Foundation of Korea (NRF)
grant (NRF-2019R1A2C2010885). We are grateful to John Hillier for making the
CMFGEN code publicly available and for the helps he gave us at the beginning of
this project. 
\end{acknowledgements}

\bibliography{ref}{}
\bibliographystyle{aasjournal}

\end{enumerate}

\appendix

\section{Optical magnitudes of SN I\texorpdfstring{\MakeLowercase{B}}{b}/I\texorpdfstring{\MakeLowercase{C}}{c} progenitors with various mass-loss rates}\label{subsec:Appendix1}

\startlongtable
\begin{deluxetable*}{l|cccc|cccc}
\tablenum{5}
\tablecaption{Optical magnitudes of SN Ib/Ic progenitors with various mass-loss rates\label{tab:Appendix1}}
\tablehead{
\colhead{}  
& \multicolumn{4}{|c}{$HST$/WFPC2} & \multicolumn{4}{|c}{$HST$/WFC3}\\ 
\cline{2-5} \cline{6-9} 
\colhead{Name}  
& \multicolumn{1}{|c}{$M_{F450W}$} & \colhead{$M_{F555W}$} & \colhead{$M_{F606W}$} & \colhead{$M_{F814W}$} 
& \multicolumn{1}{|c}{$M_{F438W}$} & \colhead{$M_{F555W}$} & \colhead{$M_{F606W}$} & \colhead{$M_{F814W}$}
\\
\colhead{}
& \multicolumn{1}{|c}{(mag)} & \colhead{(mag)} & \colhead{(mag)} & \colhead{(mag)} 
& \multicolumn{1}{|c}{(mag)} & \colhead{(mag)} & \colhead{(mag)} & \colhead{(mag)} 
}
\startdata
\hline \hline
\multicolumn{9}{c}{\emph{$\dot{M}_\mathrm{fid}\times10.0$}} \\
\hline
HE2.91 & -5.18 & -5.15 & -5.18 & -5.20 & -5.15 & -5.13 & -5.15 & -5.18\\
HE2.97 & -5.42 & -5.33 & -5.35 & -5.32 & -5.36 & -5.28 & -5.28 & -5.22\\
HE4.09 & -5.77 & -5.75 & -5.83 & -5.81 & -5.37 & -5.41 & -5.47 & -5.58\\
HE5.05 & -6.18 & -6.16 & -6.24 & -6.22 & -5.76 & -5.80 & -5.87 & -5.99\\
CO5.18 & -6.59 & -6.43 & -6.47 & -6.54 & -6.07 & -6.06 & -6.08 & -6.14\\
CO5.50 & -6.61 & -6.39 & -6.45 & -6.71 & -5.90 & -5.86 & -5.88 & -5.92\\
CO6.17 & -6.59 & -6.37 & -6.41 & -6.68 & -5.89 & -5.85 & -5.86 & -5.89\\
CO7.50 & -6.51 & -6.31 & -6.31 & -6.52 & -5.79 & -5.75 & -5.76 & -5.77\\
CO9.09 & -6.96 & -6.95 & -6.85 & -6.78 & -6.08 & -6.05 & -6.07 & -6.11\\
\hline
CO2.16 & -4.50 & -4.28 & -4.30 & -4.34 & -3.95 & -3.91 & -3.92 & -3.94\\
CO3.93 & -4.28 & -4.19 & -4.24 & -4.33 & -3.86 & -3.89 & -3.94 & -4.04\\
\hline \hline
\multicolumn{9}{c}{\emph{$\dot{M}_\mathrm{fid}\times5.0$}} \\
\hline
HE2.91 & -5.23 & -5.18 & -5.19 & -5.17 & -5.22 & -5.18 & -5.18 & -5.16\\
HE2.97 & -5.39 & -5.27 & -5.27 & -5.18 & -5.37 & -5.26 & -5.23 & -5.12\\
HE4.09 & -5.31 & -5.22 & -5.24 & -5.16 & -4.99 & -4.95 & -4.97 & -4.99\\
HE5.05 & -5.67 & -5.58 & -5.61 & -5.55 & -5.27 & -5.28 & -5.32 & -5.37\\
CO5.18 & -6.12 & -5.94 & -6.03 & -6.25 & -5.51 & -5.49 & -5.51 & -5.56\\
CO5.50 & -5.83 & -5.62 & -5.65 & -5.91 & -5.14 & -5.10 & -5.11 & -5.14\\
CO6.17 & -5.82 & -5.61 & -5,63 & -5.88 & -5.12 & -5.08 & -5.09 & -5.12\\
CO7.50 & -5.80 & -5.66 & -5.63 & -5.76 & -5.03 & -5.00 & -5.01 & -5.03\\
CO9.09 & -5.87 & -5.80 & -5.74 & -5.74 & -5.01 & -4.98 & -4.99 & -5.02\\
\hline
CO2.16 & -3.56 & -2.81 & -3.60 & -3.77 & -3.17 & -3.15 & -3.17 & -3.21\\
CO3.93 & -3.77 & -2.55 & -3.84 & -3.97 & -3.38 & -3.39 & -3.42 & -3.50\\
\hline \hline
\multicolumn{9}{c}{\emph{$\dot{M}_\mathrm{fid}\times2.0$}} \\
\hline
HE2.91 & -5.28 & -5.20 & -5.21 & -5.17 & -5.26 & -5.21 & -5.21 & -5.17\\
HE2.97 & -5.08 & -4.99 & -4.99 & -4.94 & -5.08 & -5.00 & -4.99 & -4.92\\
HE4.09 & -4.86 & -4.68 & -4.65 & -4.52 & -4.73 & -4.59 & -4.56 & -4.44\\
HE5.05 & -4.92 & -4.78 & -4.76 & -4.75 & -4.65 & -4.62 & -4.63 & -4.65\\
CO5.18 & -5.34 & -5.20 & -5.23 & -5.42 & -4.74 & -4.72 & -4.74 & -4.77\\
CO5.50 & -5.09 & -5.03 & -4.99 & -5.03 & -4.33 & -4.30 & -4.32 & -4.37\\
CO6.17 & -5.10 & -5.07 & -5.01 & -4.98 & -4.30 & -4.28 & -4.30 & -4.34\\
CO7.50 & -5.10 & -5.08 & -4.92 & -4.85 & -4.17 & -4.15 & -4.17 & -4.22\\
CO9.09 & -5.12 & -5.02 & -4.80 & -4.78 & -4.13 & -4.11 & -4.13 & -4.19\\
\hline
CO2.16 & -3.22 & -3.16 & -3.13 & -3.14 & -2.56 & -2.54 & -2.55 & -2.58\\
CO3.93 & -3.33 & -3.18 & -3.22 & -3.42 & -2.80 & -2.78 & -2.80 & -2.83\\
\hline \hline
\multicolumn{9}{c}{\emph{$\dot{M}_\mathrm{fid}\times1.0$}} \\
\hline
HE2.91 & -5.26 & -5.19 & -5.19 & -5.16 & -5.25 & -5.20 & -5.20 & -5.15\\
HE2.97 & -5.11 & -5.01 & -5.00 & -4.92 & -5.12 & -5.03 & -5.01 & -4.91\\
HE4.09 & -4.73 & -4.54 & -4.48 & -4.31 & -4.67 & -4.50 & -4.46 & -4.26\\
HE5.05 & -4.56 & -4.39 & -4.34 & -4.29 & -4.37 & -4.27 & -4.25 & -4.20\\
CO5.18 & -4.83 & -4.84 & -4.76 & -4.74 & -4.13 & -4.12 & -4.15 & -4.21\\
CO5.50 & -4.54 & -4.42 & -4.24 & -4.27 & -3.69 & -3.68 & -3.70 & -3.77\\
CO6.17 & -4.59 & -4.46 & -4.22 & -4.22 & -3.67 & -3.66 & -3.68 & -3.75\\
CO7.50 & -4.53 & -4.52 & -4.25 & -4.11 & -3.51 & -3.51 & -3.54 & -3.63\\
CO9.09 & -4.42 & -4.39 & -4.10 & -3.97 & -3.35 & -3.34 & -3.37 & -3.45\\
\hline
CO2.16 & -2.66 & -2.55 & -2.40 & -2.45 & -1.92 & -1.91 & -1.94 & -2.00\\
CO3.93 & -2.83 & -2.81 & -2.78 & -2.78 & -2.22 & -2.21 & -2.24 & -2.30\\
\hline \hline
\multicolumn{9}{c}{\emph{$\dot{M}_\mathrm{fid}\times0.5$}} \\
\hline
HE2.91 & -5.29 & -5.22 & -5.22 & -5.17 & -5.29 & -5.23 & -5.23 & -5.17\\
HE2.97 & -5.05 & -4.95 & -4.94 & -4.88 & -5.05 & -4.97 & -4.96 & -4.87\\
HE4.09 & -4.89 & -4.68 & -4.60 & -4.41 & -4.85 & -4.66 & -4.61 & -4.37\\
HE5.05 & -4.39 & -4.18 & -4.12 & -4.00 & -4.24 & -4.09 & -4.05 & -3.91\\
CO5.18 & -4.33 & -4.20 & -3.94 & -3.95 & -34.6 & -3.45 & -3.49 & -3.57\\
CO5.50 & -3.84 & -3.89 & -3.68 & -3.53 & -3.01 & -2.99 & -3.02 & -3.10\\
CO6.17 & -3.77 & -4.03 & -3.86 & -3.54 & -2.98 & -2.97 & -3.00 & -3.10\\
CO7.50 & -3.31 & -3.38 & -3.28 & -3.30 & -2.80 & -2.81 & -2.85 & -2.98\\
CO9.09 & -3.24 & -3.17 & -3.08 & -3.19 & -2.72 & -2.73 & -2.77 & -2.90\\
\hline
CO2.16 & -1.91 & -2.05 & -1.89 & -1.73 & -1.20 & -1.20 & -1.23 & -1.32\\
CO3.93 & -2.33 & -2.19 & -2.01 & -2.04 & -1.65 & -1.63 & -1.65 & -1.72\\
\hline \hline
\multicolumn{9}{c}{\emph{$\dot{M}_\mathrm{fid}\times0.1$}} \\
\hline
HE2.91 & -5.28 & -5.21 & -5.20 & -5.16 & -5.28 & -5.22 & -5.21 & -5.16\\
HE2.97 & -5.06 & -4.96 & -4.95 & -4.87 & -5.06 & -4.98 & -4.96 & -4.87\\
HE4.09 & -4.86 & -4.64 & -4.57 & -4.34 & -4.87 & -4.66 & -4.60 & -4.31\\
HE5.05 & -4.34 & -4.10 & -4.03 & -3.79 & -4.33 & -4.11 & -4.04 & -3.74\\
CO5.18 & -2.68 & -2.92 & -2.08 & -2.35 & -2.37 & -2.21 & -2.17 & -2.06\\
CO5.50 & -2.16 & -2.01 & -1.91 & -1.85 & -1.91 & -1.75 & -1.72 & -1.63\\
CO6.17 & -2.04 & -1.85 & -1.75 & -1.75 & -1.71 & -1.58 & -1.56 & -1.52\\
CO7.50 & -1.57 & -1.42 & -1.36 & -1.39 & -1.37 & -1.27 & -1.27 & -1.27\\
CO9.09 & -1.43 & -1.30 & -1.25 & -1.26 & -1.31 & -1.20 & -1.20 & -1.19\\
\hline
CO2.16 &  0.52 &  0.57 &  0.61 &  0.51 &  0.73 &  0.75 &  0.72 &  0.61\\
CO3.93 & -1.11 & -1.31 & -1.19 & -0.65 & -0.87 & -0.67 & -0.62 & -0.43\\
\enddata
\tablecomments{The meaning of each model name is same as that in Table \ref{tab:input}. The absolute magnitudes in $HST$/WFPC2 and $HST$/WFC3 filters are presented under the various mass-loss rate assumptions. $\dot{M}_\mathrm{fid}$ denotes the fiducial mass-loss rate in this study. 
}
\end{deluxetable*}

\clearpage

\section{Normalized optical spectra of fiducial models}\label{subsec:Appendix2}

\begin{figure}[thb!]
\includegraphics[width=\textwidth]{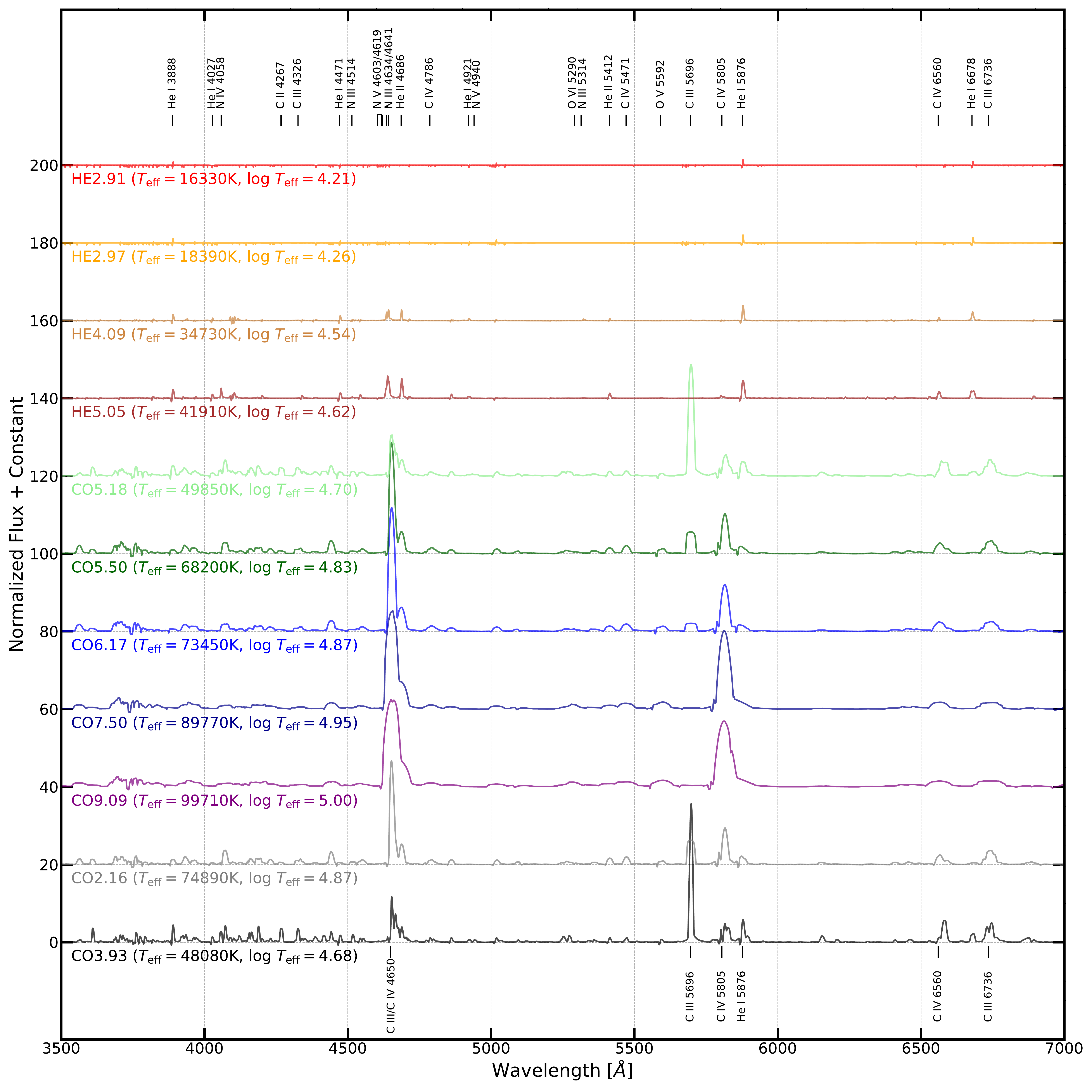}
\centering
\caption{Normalized spectra of fiducial SN Ib/Ic progenitor models in the wavelength
ranges $3500\text{\AA} \leqslant \lambda \leqslant 7000\text{\AA}$, which correspond to 
the $F435W$ and $F555W$ filter range.\label{fig:Appendix_normalized_spectra1}} 
\end{figure}

\begin{figure*}[thb!]
\includegraphics[width=\textwidth]{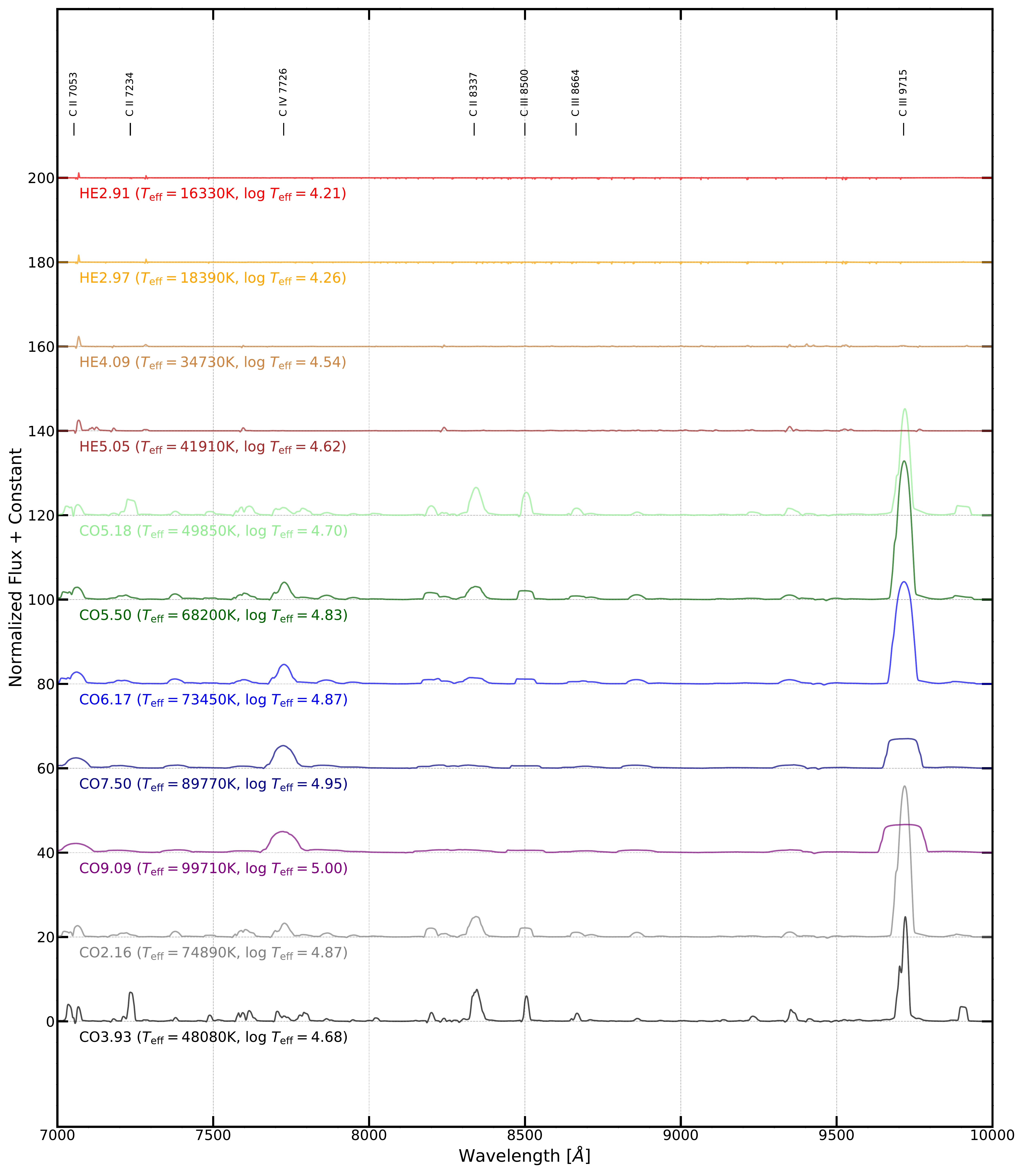}
\centering
\caption{Normalized spectra of fiducial SN Ib/Ic progenitor models in the wavelength
ranges $7000\text{\AA} \leqslant \lambda \leqslant 10000\text{\AA}$, which correspond to the  
$F814W$ filter range.\label{fig:Appendix_normalized_spectra2}} 
\end{figure*}

\clearpage

\section{Change of spectra according to various mass-loss rates}\label{subsec:Appendix3}

\begin{figure*}[thb!]
\includegraphics[width=\textwidth]{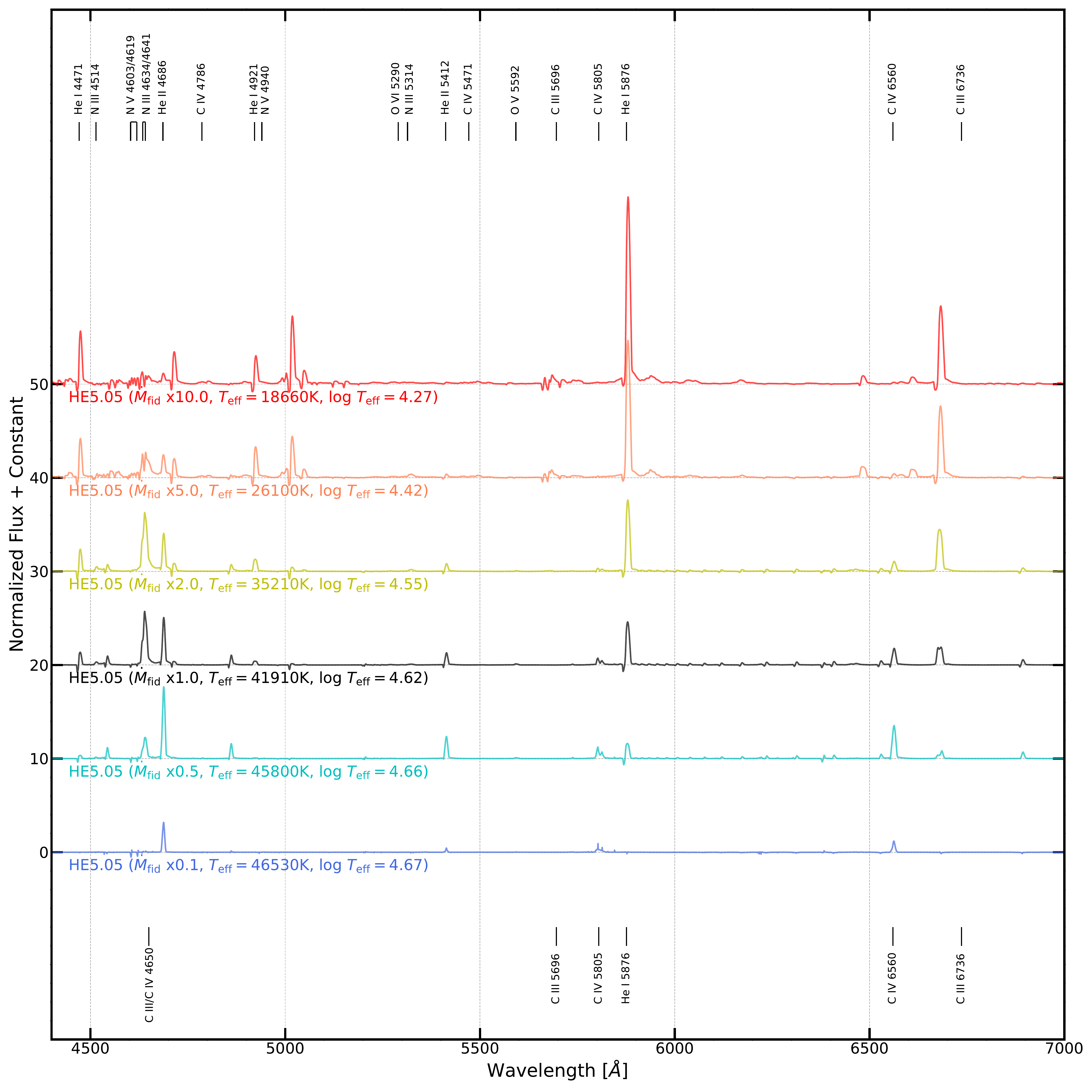}
\centering
\caption{Normalized spectra of HE5.05 model with various mass-loss rates. Spectra are presented in the wavelength
ranges $4400~\text{\AA} \leqslant \lambda \leqslant 7000~\text{\AA}$, which correspond to the wavelength range of the $F555W$ filter. \label{fig:Appendix_Ib_Mdot_normalized_spectra1}} 
\end{figure*}

\begin{figure*}[thb!]
\includegraphics[width=\textwidth]{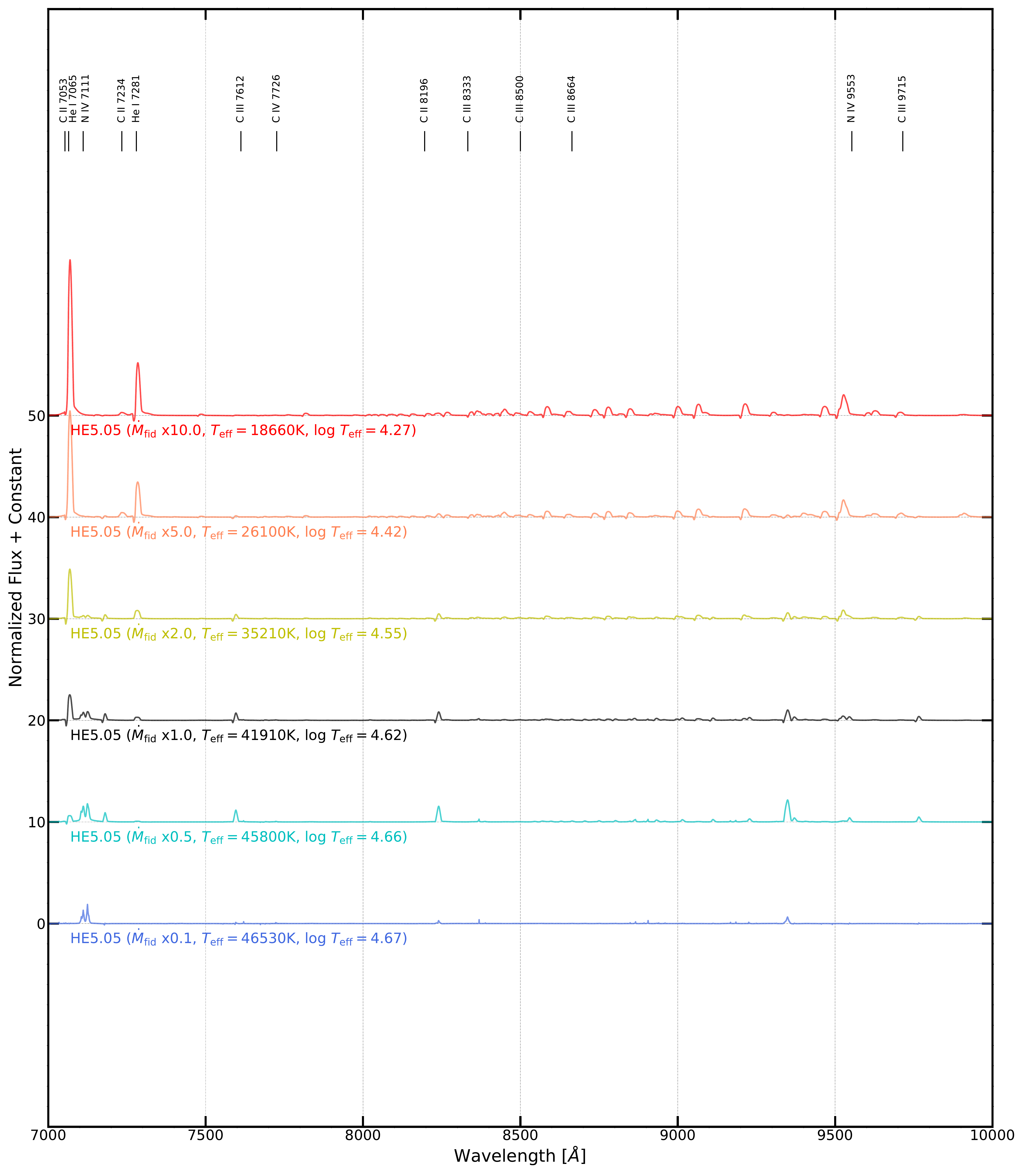}
\centering
\caption{Normalized spectra of HE5.05 model with various mass-loss rates. Spectra are presented in the wavelength
ranges $7000~\text{\AA} \leqslant \lambda \leqslant 10000~\text{\AA}$, which correspond to the wavelength range of the $F814W$ filter. \label{fig:Appendix_Ib_Mdot_normalized_spectra2}} 
\end{figure*}

\begin{figure*}[thb!]
\includegraphics[width=\textwidth]{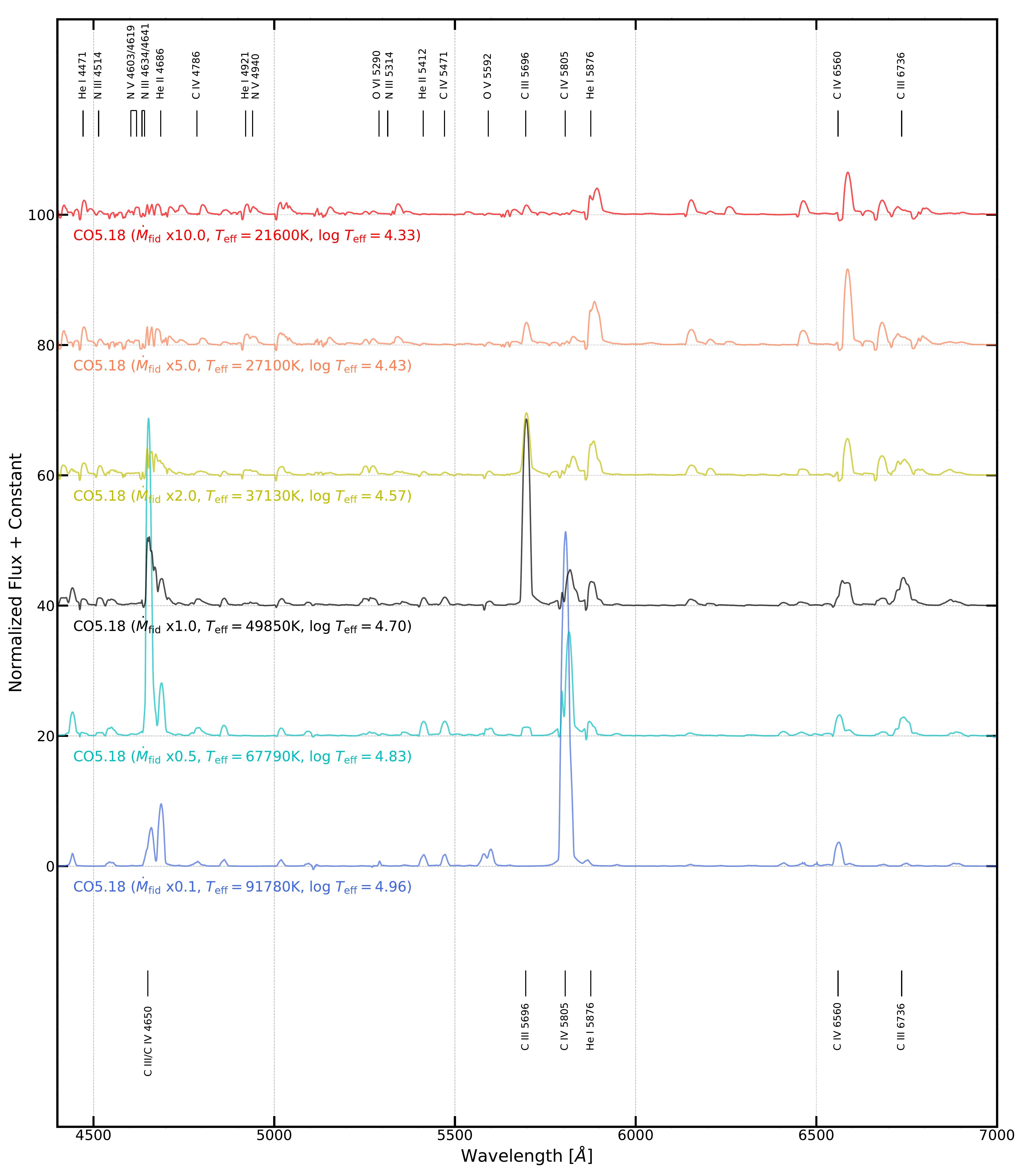}
\centering
\caption{Normalized spectra of CO5.18 model with various mass-loss rates. Spectra are presented in the wavelength
ranges $4400~\text{\AA} \leqslant \lambda \leqslant 7000~\text{\AA}$, which correspond to the wavelength range of the $F555W$ filter. \label{fig:Appendix_Ic_Mdot_normalized_spectra1}} 
\end{figure*}

\begin{figure*}[thb!]
\includegraphics[width=\textwidth]{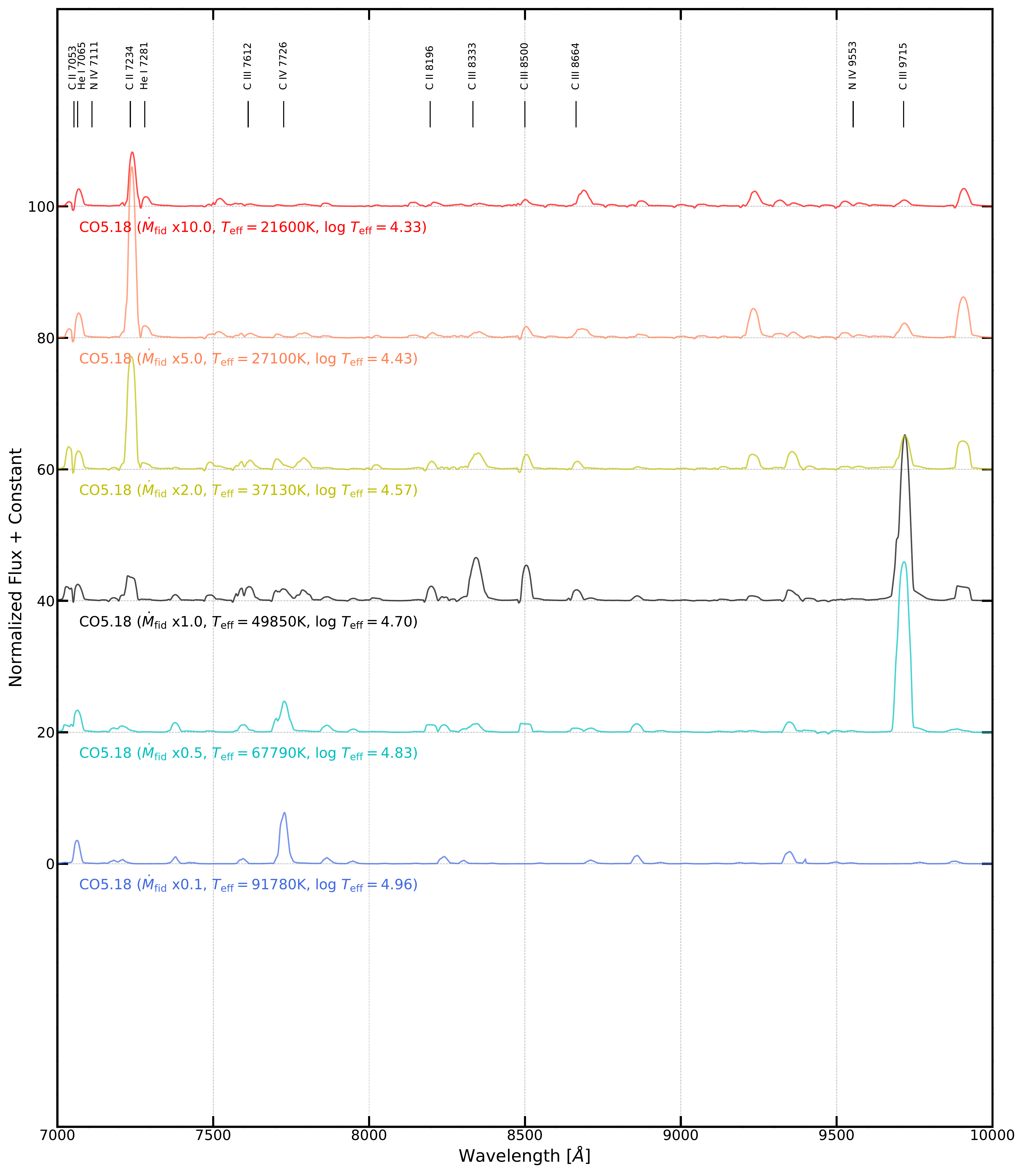}
\centering
\caption{Normalized spectra of CO5.18 model with various mass-loss rates. Spectra are presented in the wavelength
ranges $7000~\text{\AA} \leqslant \lambda \leqslant 10000~\text{\AA}$, which correspond to the wavelength range of the $F814W$ filter. \label{fig:Appendix_Ic_Mdot_normalized_spectra2}} 
\end{figure*}

\end{document}